\documentclass[11pt]{cernrep}
\usepackage{graphicx}
\usepackage{cite,./mcite}
\usepackage{wrapfig,rotating}
\usepackage[pagewise]{lineno}

\def\gsim{\raise0.3ex\hbox{$>$\kern-0.75em\raise-1.1ex\hbox{$\sim$}}}
\def\lsim{\raise0.3ex\hbox{$<$\kern-0.75em\raise-1.1ex\hbox{$\sim$}}}

\begin{document}
\title{Direct photon production at HERA, the Tevatron and the LHC}

\author{R.~E.~Blair $^a$, S.~Chekanov  $^a$, G.~Heinrich $^b$, A.Lipatov $^c$ and  N.Zotov $^c$ }
\date{Sep 2008}
\documentlabel{ANL-HEP-CP-08-52\\IPPP/08/64\\DCPT/08/128 }

\institute{
\vspace{0.5cm}$^a$ HEP Division, Argonne National Laboratory,
9700 S.Cass Avenue,
Argonne, IL 60439,
USA \\
$^b$ Institute for Particle Physics Phenomenology,
        Department of Physics,\\
        $\>\>\>$University of Durham, Durham, DH1 3LE, UK \\ 
$^c$ SINP, Moscow State University, 119991 Moscow,  Russia 
}

% put line numbers
% \linenumbers

\maketitle

\begin{abstract}
We review several most recent prompt-photon measurements at HERA and the Tevatron
and discuss their implication for future measurements at the LHC.
A comparison to Monte Carlo models, as well as to NLO
QCD predictions based on the standard DGLAP and
the $k_T$-factorization approaches is discussed.
Effects from renormalization and factorization scale uncertainties, as well
as uncertainties on the gluon density distribution inside a proton are discussed. 
\end{abstract}

\section{Introduction}

Events with an isolated photon are an important tool
to study hard interaction processes since such photons
emerge without the hadronization phase.
In particular, final states of $ep$ and $pp$ collisions with a
prompt photon together with a jet are more directly sensitive
to the underlying QCD process than inclusive prompt photon measurements.

\begin{wrapfigure}{r}{0.35\columnwidth}
\centerline{\includegraphics[width=0.25\columnwidth]{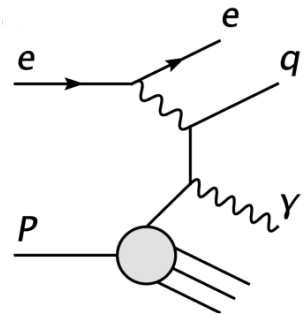}}
\caption{\label{fig:compton} Lowest-order diagram (Compton scattering)
for $\gamma$+jet events in $ep$ collisions}.
\end{wrapfigure}

\begin{figure}[ht]
\begin{minipage}[b]{0.3\linewidth}
\centering
\includegraphics[scale=0.5]{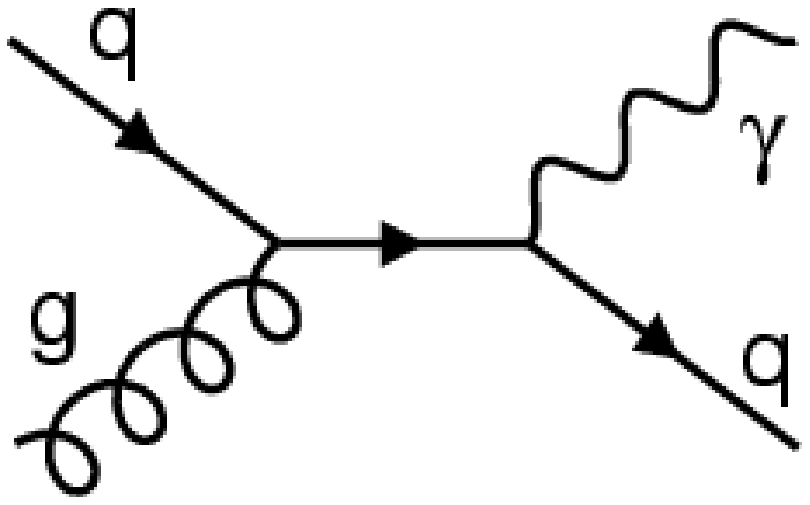}
\end{minipage}
\hspace{0.6cm}
\begin{minipage}[b]{0.3\linewidth}
\centering
\includegraphics[scale=0.25]{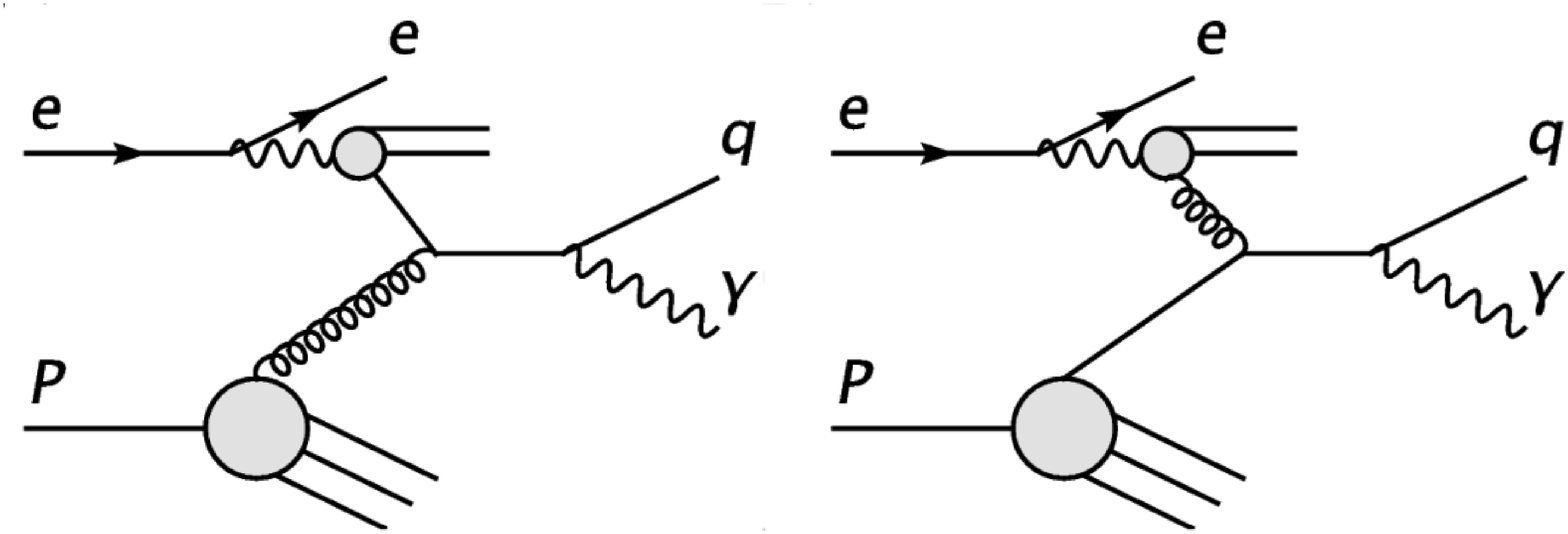}
\end{minipage}
\caption{
The dominant diagram for prompt-photon events in $pp$ collisions (left figure) 
compared to two resolved photon diagrams in $ep$ photoproduction 
 (see  Section~\ref{NLO} for more details).  
}
\label{fig:ppvsep}
\end{figure}

The results on prompt-photon production 
provided by HERA are important for the interpretation of the LHC data.
Unlike $pp$ collisions, 
$ep$ collisions
involve a point-like incoming lepton, which
leads to some simplification
in the description of the prompt-photon production in $ep$ compared to $pp$.
At HERA, the quark content of the proton is probed through the elastic scattering of a
photon by a quark, $\gamma q\to\gamma q$ (see Fig.~\ref{fig:compton}).
Such QED events are significantly
simpler than lowest-order  “compton-like” $qg\to q\gamma$ events which are
dominant in $pp$ collisions (see Fig.~\ref{fig:ppvsep}, left figure). 
The latter process has direct sensitivity to the strong coupling constant 
and requires  much better understanding of the gluon structure function inside 
both incoming protons than for the lowest-order diagram in $ep$ collisions.  

Despite  the 
difference between $ep$ and $pp$ collisions concerning certain lowest-order diagrams, 
a large class of partonic contributions are similar between
$ep$ and $pp$ collisions, due to the hadronic nature of the resolved photon.
In particular, a contribution to prompt-photon
events from the $gq \to q\gamma$ process in photoproduction, in which the photon 
displays its hadronic structure 
\cite{Owens:1986mp,pr:d52:58,pr:d64:14017,ejp:c21:303},
leads to significant
sensitivity to the gluon structure function as is the case in $pp$ collisions 
(see Fig.~\ref{fig:ppvsep}, the two figures on the right).   
Therefore, analysis of HERA data can make a bridge between 
a better understood $ep$ case 
and the less understood $pp$ case, since 
apart from the convolution with different structure functions, photoproduction 
diagrams $ep$ collisions involving a resolved photon 
are essentially the same as  diagrams in $pp$ collisions. 

Prompt-photon events in $ep$ collisions 
can constrain both quark and gluon
parton densities (PDFs).
In addition, differences between
collinear factorization and $k_T$ factorization 
in the description of the underlying
hard subprocesses
can be studied in detail. This is important not only
for a better understanding of QCD dynamics, but also has direct implications 
for searches of exotic physics at the LHC, in which prompt-photon production is
the main background. 
A number of QCD predictions
\cite{pr:d52:58, pr:d64:14017, ejp:c21:303,ejp:c34:191,Lipatov:2005tz}
can be confronted with the data and some of them will be described in more detail below.

\section{Photoproduction of prompt photons at NLO} 
\label{NLO}

\begin{center}
\begin{figure}[htb]
\begin{picture}(100,210)(-100,-115)
%\SetScale{0.7}
%\SetWidth{0.5}
\put(0,0){\includegraphics[scale=0.75]{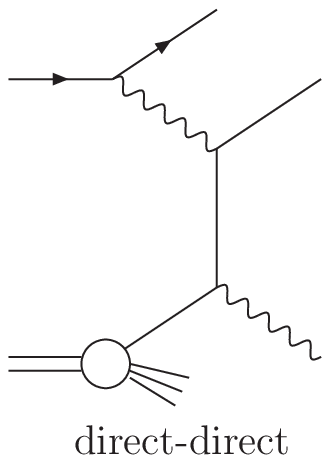}}
\put(100,0){\includegraphics[scale=0.75]{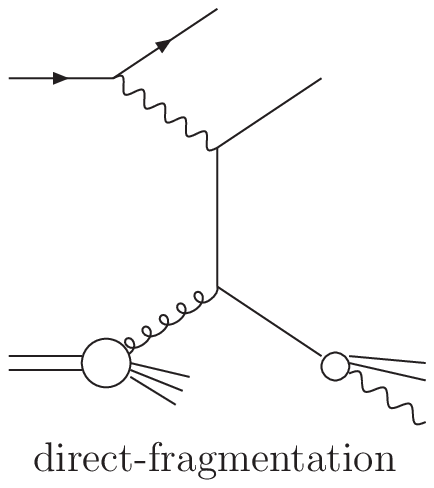}}
\put(0,-125){\includegraphics[scale=0.75]{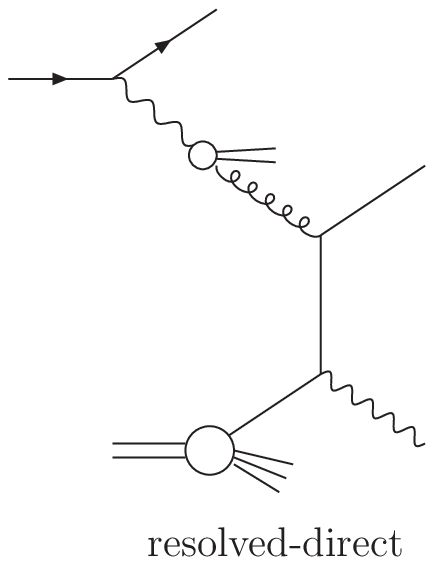}}
\put(100,-125){\includegraphics[scale=0.75]{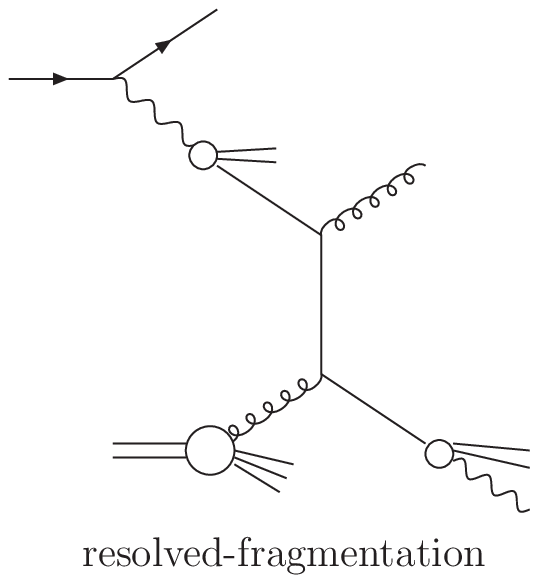}}
\end{picture}
\caption{Examples of contributing subprocesses at leading order to each of the four categories in $ep$ collisions.}\label{fig:cats}
\end{figure}
\end{center}

In the photoproduction $ep$ scattering processes, the electron 
is scattered at small angles, emitting a quasi-real photon which 
scatters with the proton. The spectrum of these photons can be described by the 
Weizs\"acker-Williams approxi\-mation\,\cite{vonWeizsacker:1934sx,*Williams:1934ad}. 
The photons will take part in the hard interaction either directly, 
or through their ``partonic" content, in which case a parton stemming from 
the {\em resolved} photon participates in the hard subprocess.
Similarly, a photon in the final state can either originate directly from 
the hard interaction or from the fragmentation of a parton.
Therefore, one can distinguish four categories of subprocesses:
1) direct direct, 2) direct fragmentation,  3) resolved direct and 4) resolved fragmentation. 
Examples of leading order diagrams of each class are shown in Fig.~\ref{fig:cats}.
Beyond leading order, this distinction becomes ambiguous.
For example, the NLO corrections to the direct part involve 
final state collinear quark-photon pairs  which lead to divergences which 
are absorbed into the fragmentation function, such that only the 
sum of these contributions has a physical meaning.
The complete NLO corrections to all four parts have been calculated in 
\cite{ejp:c21:303} for inclusive prompt photons and in \cite{Fontannaz:2001nq} 
for photon plus jet final states. 
A public program {\small EPHOX}, written as a partonic event generator, is available from \cite{phox}. The NLO corrections to the direct-direct part also have been calculated in \cite{pr:d64:14017,gamjet}  for the inclusive and photon plus jet final state.

The $\gamma$-$p$ scattering processes are of special interest 
since they are sensitive to both the partonic structure of the photon 
as well as of the proton. 
They offer the possibility  to constrain the 
(presently poorly known) gluon distributions in the photon, 
since in a certain kinematic region the subprocess 
$q g\to \gamma q$, where the gluon is stemming from a resolved photon,
dominates\,\cite{ejp:c34:191}. 

\begin{figure}[ht]
\begin{minipage}[b]{0.47\linewidth}
\centering
\includegraphics[scale=0.47]{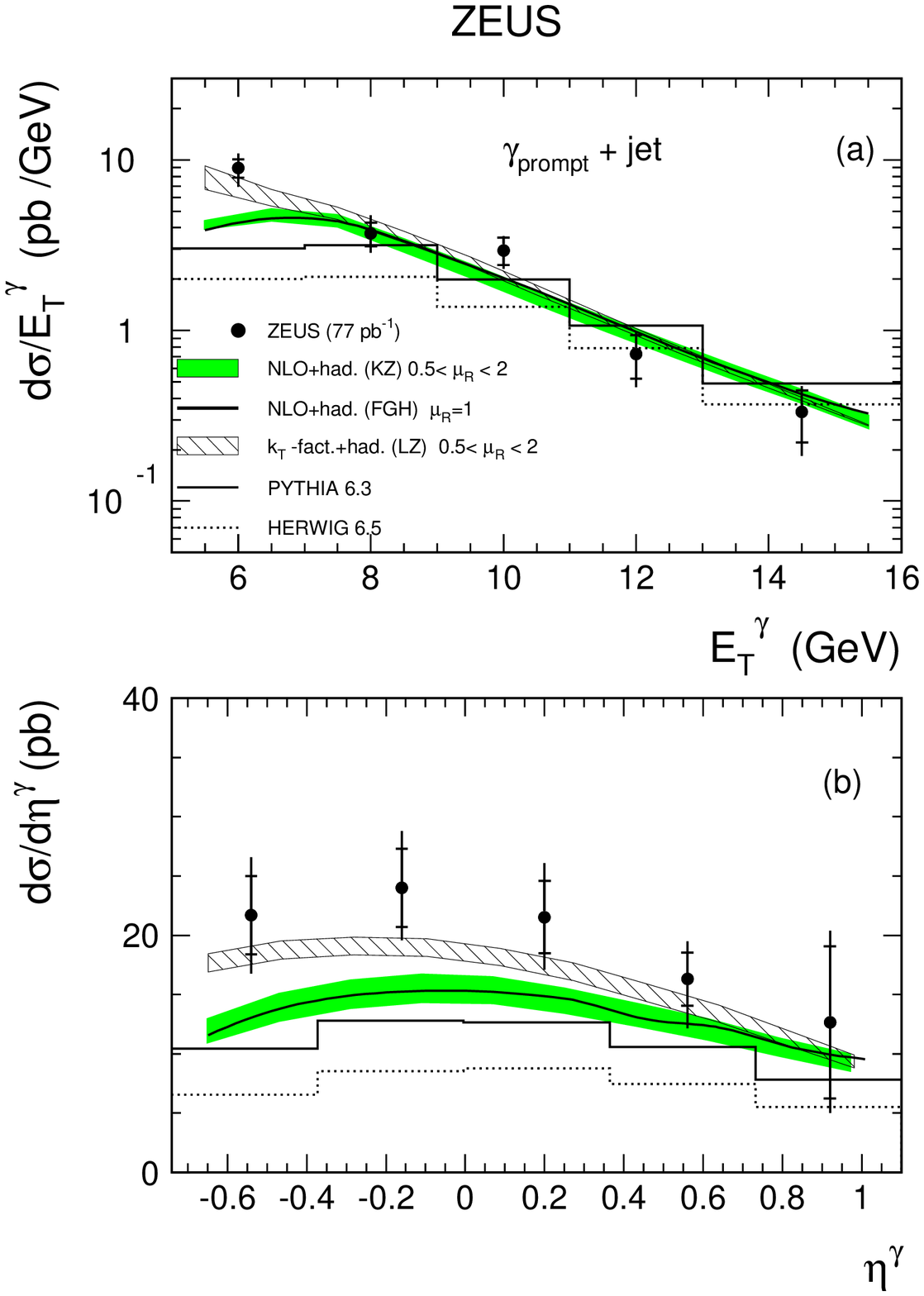}
% \caption{default}
% \label{fig:figure1}
\end{minipage}
\hspace{0.5cm}
\begin{minipage}[b]{0.47\linewidth}
\centering
\includegraphics[scale=0.47]{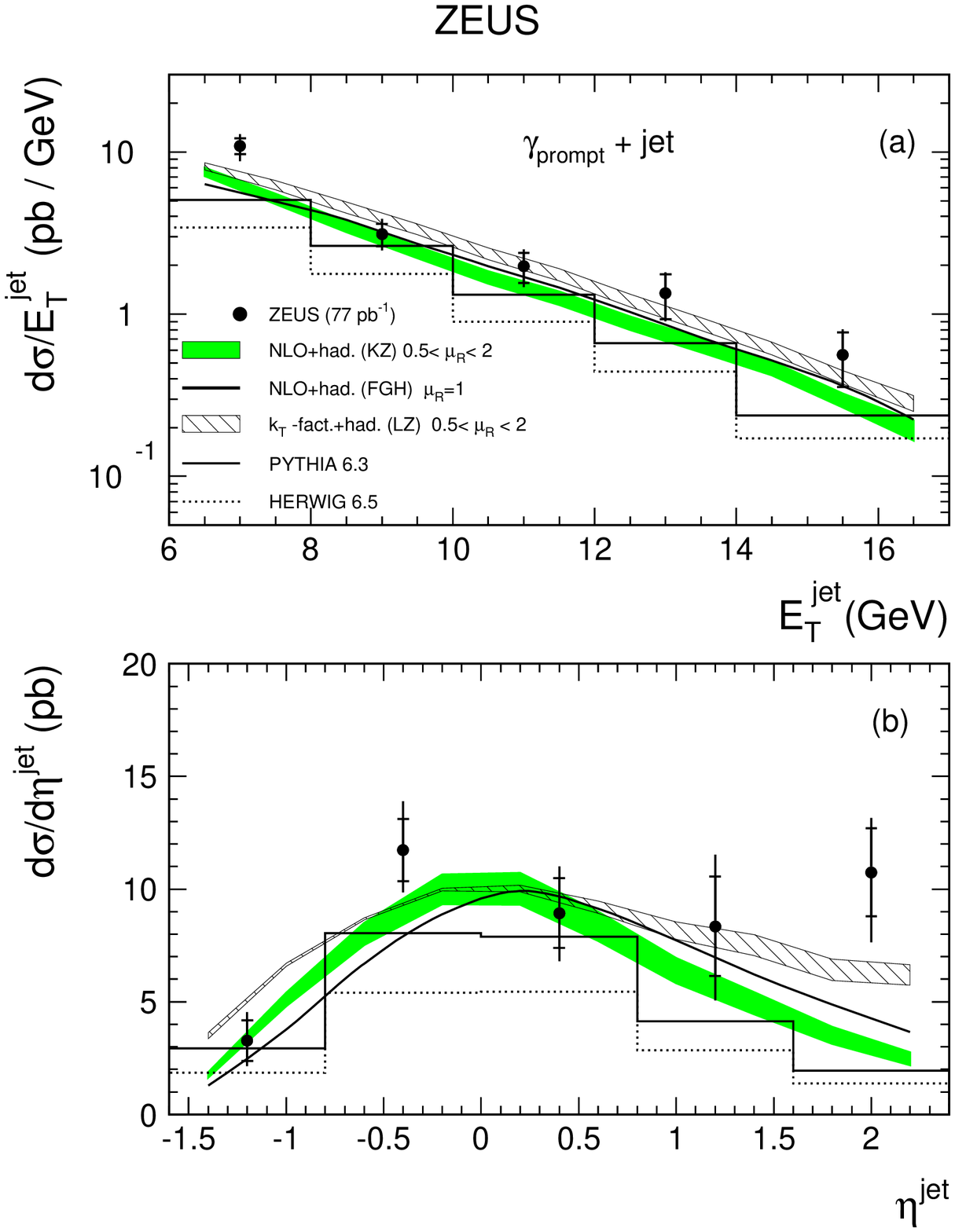}
% \caption{default}
% \label{fig:figure2}
\end{minipage}
\caption{
The differential $\gamma$+jet cross sections
as functions of $E_{T}$ and $\eta$  of  the prompt
photon and the jet.
The data are compared to QCD  calculations
and Monte Carlo models as described in the text.
The shaded bands correspond to a typical
scale uncertainty which was obtained by changing the renormalization and
factorization scales simultaneously by a factor of 0.5 and 2 respectively.
}
\label{fig:zeus}
\end{figure}

Working within the framework of collinear factorization, 
i.e. assuming that the transverse momenta of the partons within the proton
can be neglected and other non-perturbative effects can be
factorized from the hard interaction at high momentum transfers,
the cross section for $e p\to\gamma X$ can symbo\-li\-cally 
be written as a convolution of the parton densities for the incident particles
(respectively fragmentation function for an outgoing parton  fragmenting into a
photon) with the partonic cross section $\hat \sigma$:
\begin{eqnarray}
d\sigma^{ep\to\gamma X}(P_p,P_e,P_{\gamma})=
\sum_{a,b,c}\int dx_e\int d x_p
\int
dz\, F_{a/e}(x_e,M)F_{b/p}(x_p,M_p)D_{\gamma/c}(z,M_F)\nonumber\\
d\hat\sigma^{ab\to c
X}(x_pP_p,x_eP_e,P_{\gamma}/z,\mu,M,M_p,M_F)\;,
\label{dsigma}
\end{eqnarray}
where $M,M_p$ are the initial state factorization scales, $M_F$ the 
final state factorization scale, $\mu$ the
renormalization scale and $a,b,c$ run over parton types. 
In the NLO calculations shown in Fig.~\ref{fig:zeus}, 
all these scales are set equal to $p_T^\gamma$ and varied simultaneously.
The functions $F_{b/p}(x_p,M_p)$ are the parton distribution functions in the proton, 
obeying DGLAP evolution.
Note that including initial state radiation at NLO 
in the partonic calculation means that the partons 
taking part in the hard interaction can pick up a nonzero transverse 
momentum. In certain cases, this additional ``$k_T$-kick" seems to be sufficient 
to describe the data well. 
For example, a study of the effective transverse momentum $\langle k_T\rangle$ 
of partons in the proton has been made by ZEUS\,\cite{Chekanov:2001aq}. 
Comparing the shapes of normalized distributions for 
$\langle k_T\rangle$-sensitive ob\-ser\-vables to an NLO calculation, 
it was found that the data agree well with NLO QCD 
without extra intrinsic $\langle k_T\rangle$\,\cite{Fontannaz:2001nq}.

The ``resolved'' contributions are characterized by a resolved photon in the
initial state where a parton stemming from the  photon 
instead of the photon itself parti\-ci\-pates in the hard subprocess. 
In these cases, $F_{a/e}(x_e,M)$ is given by a convolution of the 
Weizs\"acker-Williams spectrum $f^e_{\gamma}(y)$ with the parton distributions 
in the photon:
\begin{eqnarray}\label{phpdf}
F_{a/e}(x_e,M)=
\int_0^1 dy \,dx_{\gamma}\,f^e_{\gamma}(y) \,
F_{a/\gamma}(x_{\gamma},M)\,\delta(x_{\gamma}y-x_e)\;.
\end{eqnarray}
The cases with ``direct" attributed to the initial state photon 
correspond to  $a=\gamma$, so $F_{a/\gamma}=\delta(1-x_\gamma)$ and
 $F_{a/e}$ in eq.~(\ref{phpdf}) collapses to 
the Weizs\"acker-Williams spectrum. 
The cases ``direct-direct" and ``resolved-direct" correspond to 
$c=\gamma$, so $D_{\gamma/c}(z,M_F)=\delta_{c\gamma}\delta(1-z)$ 
in (\ref{dsigma}), i.e. the prompt  
photon is produced directly in the hard subprocess and not from the 
fragmentation of a hard parton.

If additional jets are measured, eq.~(\ref{dsigma}) also contains a jet function, 
which defines the clustering of the final state partons other than the photon into jets.
Prompt photon production in association with a jet offers more 
possibilities to probe the underlying parton dynamics.  It allows for the definition of observables that provide information about
the longitudinal momentum fractions $x^\gamma,x^p$ carried by the particles taking part in the 
hard interaction.
The partonic $x^\gamma,x^p$  are not observable, but 
one can define the observables 
%\vspace*{-3mm}  
\begin{eqnarray}
x_{obs}^{\gamma}=
\frac{p_T^\gamma\,{\rm{e}}^{-\eta^\gamma}+p_T^{\rm{jet}}\,{\rm e}^{
-\eta^{\rm{jet}}}}{2E^{\gamma}}\;,\nonumber\\
x_{obs}^{p}=
\frac{p_T^\gamma\,{\rm{e}}^{\eta^\gamma}+p_T^{\rm{jet}}\,{\rm e}^{
\eta^{\rm{jet}}}}{2E^{p}}\;,
\label{x_gam}
\end{eqnarray}
which, for direct photons in the final state, 
coincide with the partonic $x^\gamma,x^p$ at leading order.
Unique to photoproduction processes is the 
possibility to ``switch on/off" the resolved photon by suppressing/enhancing large
$x^\gamma$. 
As $x^\gamma=1$ corresponds to direct photons in the initial state, one can 
obtain resolved photon enriched data samples by placing a cut 
$x_{obs}^\gamma\leq 0.9$. Another possibility to enhance 
or suppress the resolved photon 
component is to place cuts on $p_T$ and rapidity.
From eq.~(\ref{x_gam}) one can easily see that $x_{obs}^\gamma$ is 
small at low $p_T^{\gamma,\rm{jet}}$ values and large rapidities. 
Small $x^\gamma$-enriched data samples could be used to further 
constrain the parton distributions in the real photon, in particular 
the gluon distribution, as investigated e.g. in \cite{ejp:c34:191}. 
Similarly, one can suppress the contribution from the resolved photon 
to probe the proton at small $x^p$ by direct 
$\gamma$-$p$ interactions \cite{ejp:c34:191}. 

In order to single out the prompt photon events from the background
of secondary photons produced by the decays of light mesons,
isolation cuts have to be imposed on the photon signals in the experiment. 
A widely used isolation criterion is the following:
A photon is isolated if, inside a cone centered around the photon direction 
in the rapidity and azimuthal angle plane, the amount of hadronic transverse 
energy
$E_T^{had}$ deposited is smaller than some value $E_{T,\rm{max}}$\,: 
%fixed by theexperiment:
\begin{equation}\label{criterion}
%\left.
\begin{array}{rcc} 
\mbox{for }\;\left(  \eta - \eta_{\gamma} \right)^{2} &+&  \left(  \phi - \phi_{\gamma} \right)^{2}  
 \leq   R,\\
E_T^{\rm{had}} & \leq & E_{T,\rm{max}}\,.
\end{array}
%\right\} \nonumber
\end{equation}
HERA experiments mostly used  
$E_{T,\rm{max}}=\epsilon \,p_T^{\gamma}$ with 
$\epsilon=0.1$ and $R$ = 1.
Isolation not only reduces the background from secondary photons, but also 
substantially  reduces the contribution from the fragmentation
of hard partons into high-$p_T$ photons. 
When comparing the result of partonic calculation to data, 
photon isolation is a delicate issue. For example, 
a part of the hadronic energy measured in the cone 
may come from the underlying event; therefore even the direct contribution 
can be cut by the isolation condition if the latter is too stringent.

\section{$k_T$-factorization approach}

A complementary description is offered by the $k_T$-factorization 
approach~\cite{sovjnp:53:657,*np:b366:135,*np:b360:3},
which relies on parton distribution functions where the $k_T$-dependence
has not been integrated out. 

In the framework of $k_T$-factorization approach the treatment
of $k_T$-enhancement in the inclusive prompt photon suggests a possible modification 
of the above simple $k_T$ smearing picture.
In this approach the transverse momentum of incoming partons
is generated in the course of 
non-collinear parton evolution under control of
relevant evolution equations.
In the papers ~\cite{Lipatov:2005tz,Lipatov:2005wk}  the Kimber-Martin-Ryskin (KMR) 
formalism~\cite{Kimber:2001sc,*Watt:2003mx} 
 was applied to study the role of
the perturbative components of partonic $k_T$ in describing of the observed
$E_T$ spectrum at HERA and Tevatron.
The proper off-shell
expressions for the matrix elements of the partonic subprocesses and
% were used in \cite{Lipatov:2005wk}to analysis the Tevatron data. In addition,   
the KMR-constructed unintegrated parton densities obtained independently
were used in \cite{Lipatov:2005wk} to analyze the Tevatron data.

%%%%%%%%%%%%%%%%%%%%%%%%%%%%%%%%%%%%%%%%
\section{Comparison with HERA results}
%%%%%%%%%%%%%%%%%%%%%%%%%%%%%%%%%%%%%%%

Recently published~\cite{zeus_prph} ZEUS differential cross sections as functions of $E_{T}$ and $\eta$ for the prompt-photon
candidates and for the accompanying jets have revealed some difference with both Monte Carlo predictions and
the next-to-leading order (NLO) calculations based on the collinear factorization
and the DGLAP formalism \cite{pr:d64:14017,ejp:c21:303}, as shown in Fig.~\ref{fig:zeus}.
The data are compared to QCD  calculations
performed
by Krawczyk and Zembrzuski (KZ) \cite{pr:d64:14017},
by Fontannaz, Guillet and Heinrich (FGH)~\cite{ejp:c21:303},
by A.~Lipatov and N.~Zotov (LZ)~\cite{Lipatov:2005tz} and
and  {\sc Pythia} 6.4 \cite{Sjostrand:2006za} and {\sc Herwig} 6.5  \cite{Corcella:2000bw,*Corcella:2002jc}
Monte Carlo models.
The MC  differential cross sections do not rise as steeply
at low $E_{T}^{\gamma}$ as do the data.
It should be pointed out that no intrinsic transverse momentum of the initial-state
partons in the proton was assumed for these calculations.
The QCD calculation~\cite{Lipatov:2005tz}  based on
the $k_T$-factorization~\cite{sovjnp:53:657,*np:b366:135,*np:b360:3,*Watt:2003mx} and
the
Kimber-Martin-Ryskin (KMR) prescription \cite{Kimber:2001sc,*Watt:2003mx}
for unintegrated quark and gluon
densities,
gives the best description
of the $E_{T}$ and $\eta$ cross sections.

\begin{figure}[htb]
\centering
\includegraphics[scale=0.65]{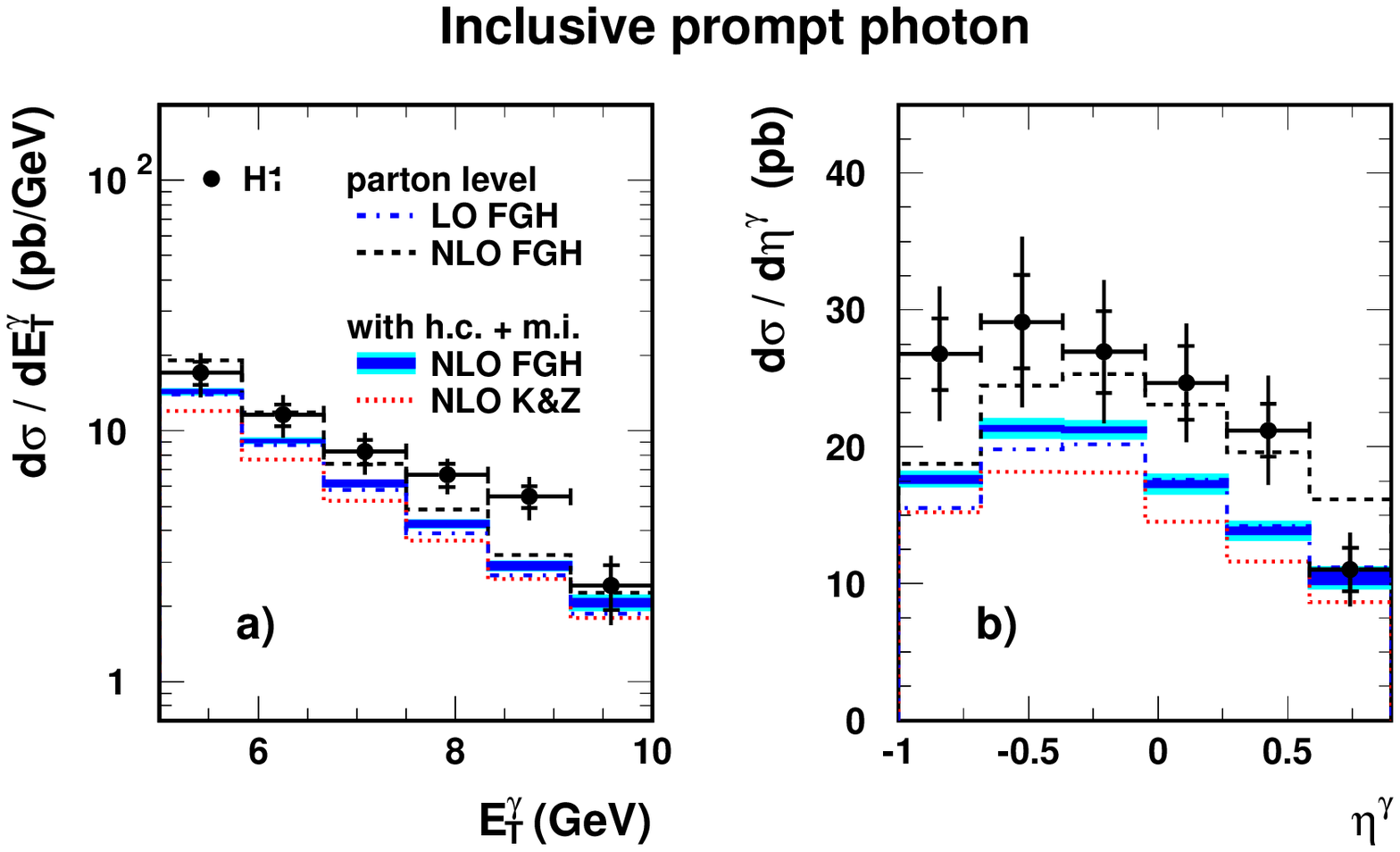}
\caption{
The differential  cross section $d\sigma/dE_T$ and $d\sigma/d\eta^\gamma$
as functions of $E_{T}^\gamma$ and $\eta^\gamma$ of the inclusive prompt
photon photoproduction calculated at $-0.7 < \eta^\gamma < 0.9$ and $0.2 < y <0.9$.
The data are compared to two different NLO calculations.}
\label{fig:H1incl}
%  \end{figure}

\vspace{0.4cm}
% \begin{figure}[ht]
\centering
\includegraphics[scale=0.65]{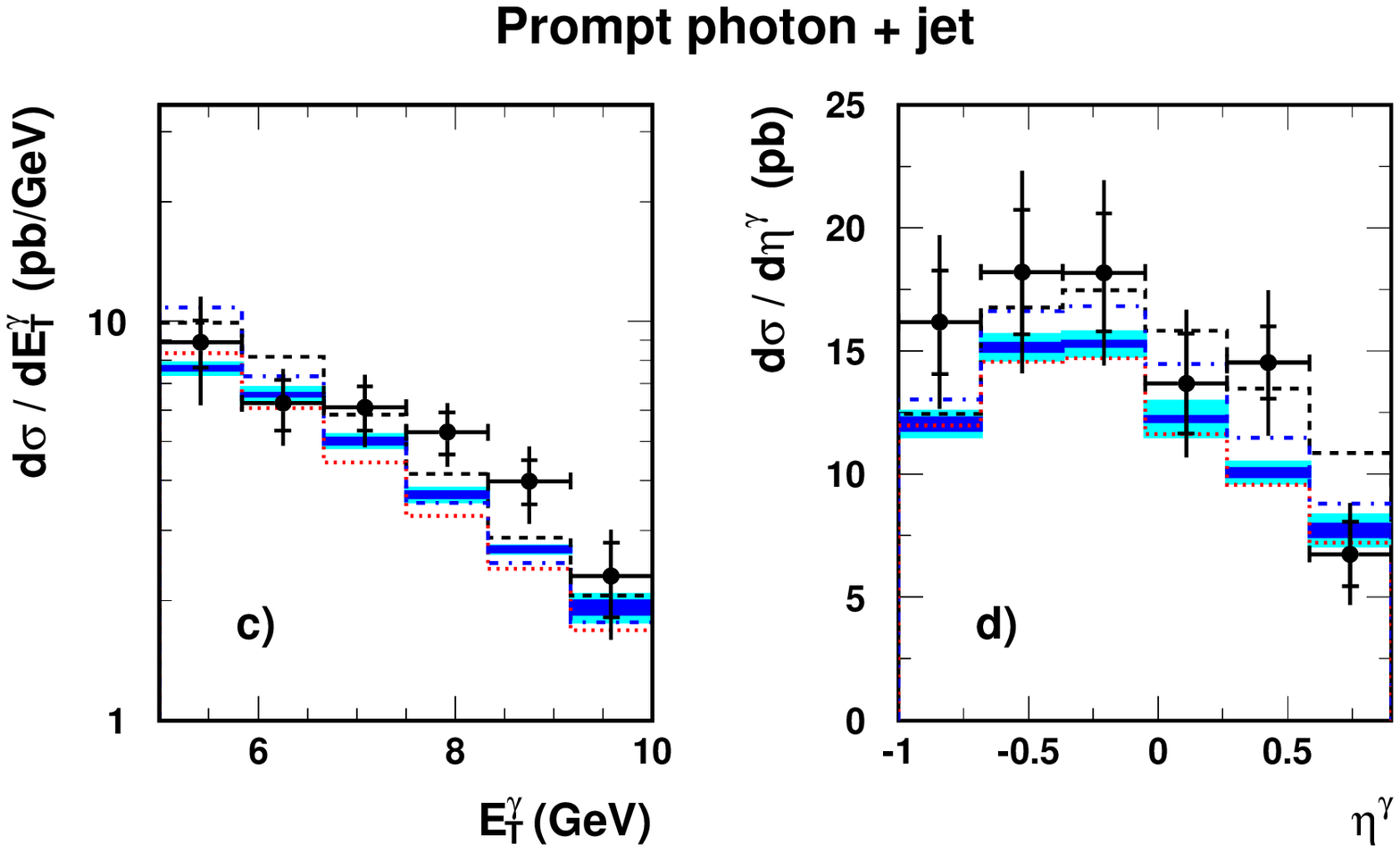}
\caption{
Same as in Fig.~\ref{fig:H1incl}, but for
$\gamma$+jet events with
the additional jet cuts:
$ - 1 < \eta^{\rm jet} < 2.3$ and $E_T^{\rm jet} > 4.5$ GeV.
}
\label{fig:H1jets}
\end{figure}

In the photon-rapidity distribution of Fig.~\ref{fig:zeus},
the  data lying above the NLO theory prediction at low
values of $\eta^\gamma$ could be explained by
the fact that in this region,  $x_{obs}^{p}$ is small,
as can be seen from Eq.~(\ref{x_gam}),
and therefore $k_T$-effects may be important.
On the other hand, this is not corroborated by the jet rapidity distribution,
which has a problem at high $\eta^{\rm{jet}}$, corresponding to
small $x_{obs}^{\gamma}$. 
Indeed, a direct measurement~\cite{zeus_prph}
of $x_{obs}^{\gamma}$ shows that the differences with NLO are mainly at low
values of the $x_{obs}^{\gamma}$ distribution.
In this region, resolved photon
events dominate, which may indicate that resolved photon remnants
could have lead to an increase in the number of jets
which have passed the experimental cuts, while these events
are not accounted for in the partonic calculation.

The inclusive prompt photon data~\cite{Aktas:2004uv,Breitweg:1999su}
lie above the NLO theory prediction in the whole rapidity range,
except for the bin of largest rapidity, where the agreement is good
after hadronization corrections, see Fig.~\ref{fig:H1incl}.

Interestingly, ZEUS investigated what happens if the
minimum transverse energy of the prompt photon is increased
to 7\,GeV, and found that in this case,
the NLO calculations are in good agreement\,\cite{zeus_prph} ,
which suggests that non-perturbative effects
may produce the discrepancy.
See \cite{Heinrich:2007mx} and references therein for more details.

The H1 experimental data in photoproduction~\cite{Aktas:2004uv} are shown
in Figs.~\ref{fig:H1incl} and \ref{fig:H1jets}.
Both  inclusive and $\gamma$+jet cross sections were compared
to the FGH NLO calculations after hadronization corrections.
The H1 data~\cite{Aktas:2004uv} referred to the kinematic region defined by
$5 < E_T^\gamma < 10$ GeV, $ - 1 < \eta^\gamma < 0.9$ and $0.2 < y < 0.7$,
which is rather similar to the ZEUS measurement shown in Fig.~\ref{fig:zeus}.
Similar to the ZEUS case,
MC predictions were found to underestimate the H1 cross sections,
while NLO QCD gives a much better description.
After taking into account
hadronization and multiple interaction effects, NLO calculations predict
somewhat smaller cross sections compared the measurements~\cite{Aktas:2004uv}.

\begin{figure}[ht]
\begin{minipage}[b]{0.48\linewidth}
\centering
\includegraphics[scale=0.7]{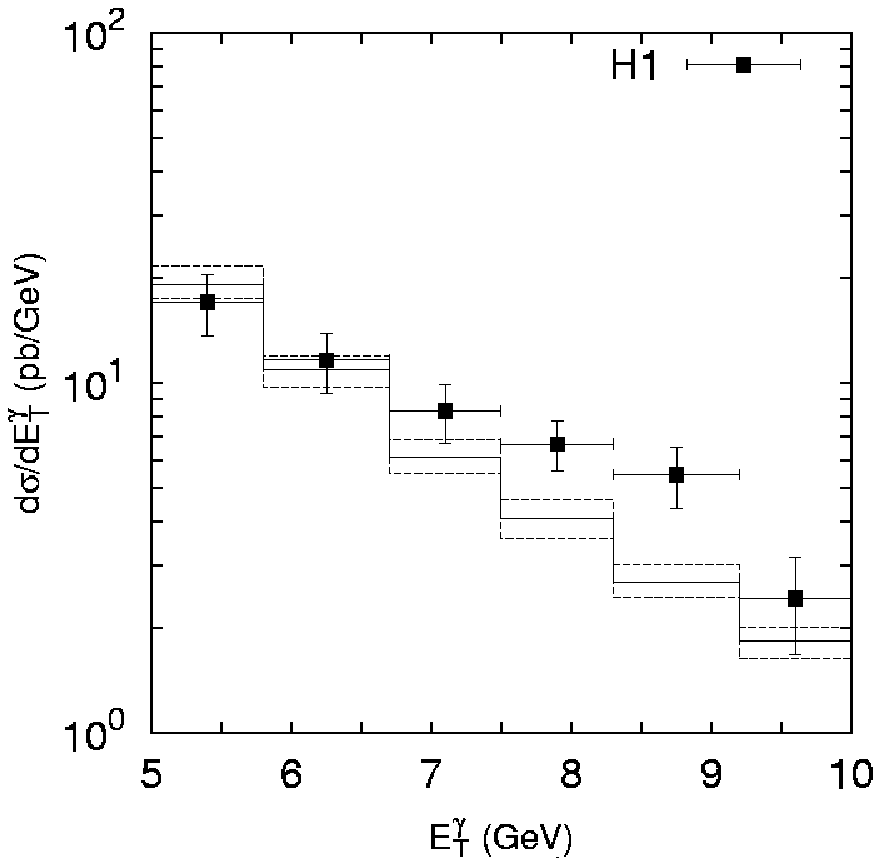}
\end{minipage}
\hspace{0.5cm}
\begin{minipage}[b]{0.48\linewidth}
\centering
\includegraphics[scale=0.7]{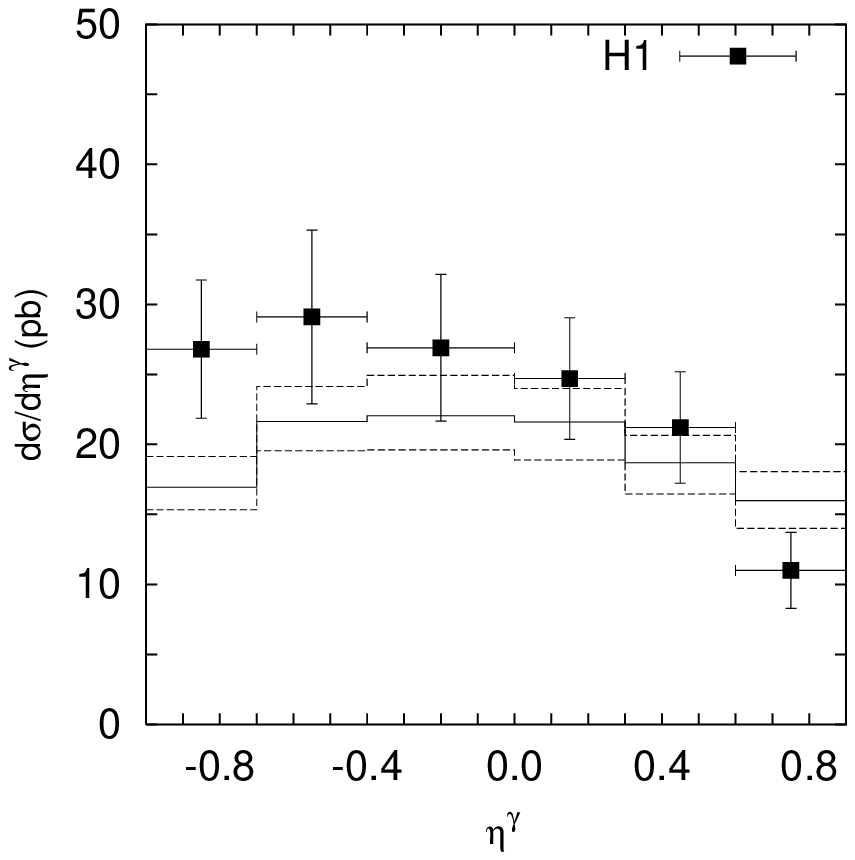}
\end{minipage}
\caption{
The differential  cross section $d\sigma/dE_T$ and $d\sigma/d\eta^\gamma$
as functions of $E_{T}^\gamma$ and $\eta^\gamma$ of the inclusive prompt
photon photoproduction calculated at $-0.7 < \eta^\gamma < 0.9$ and $0.2 < y <0.9$.
The data are compared to the $k_T-$ factorization calculations.
The  bands correspond to a typical  renormalization
scale uncertainty which was obtained by changing $\mu_R$ by a factor of 0.5 and 2.
}
\label{fig:h1cut}

\begin{minipage}[b]{0.48\linewidth}
\centering
\includegraphics[scale=0.7]{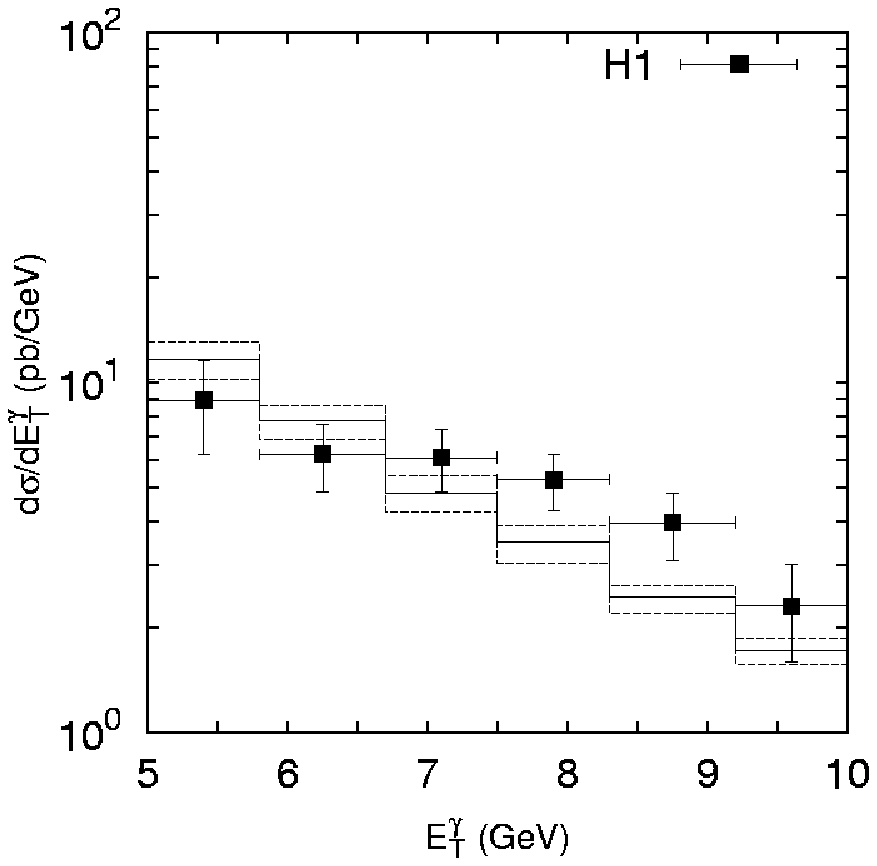}
% \caption{default}
% \label{fig:figure1}
\end{minipage}
\hspace{0.5cm}
\begin{minipage}[b]{0.48\linewidth}
\centering
\includegraphics[scale=0.7]{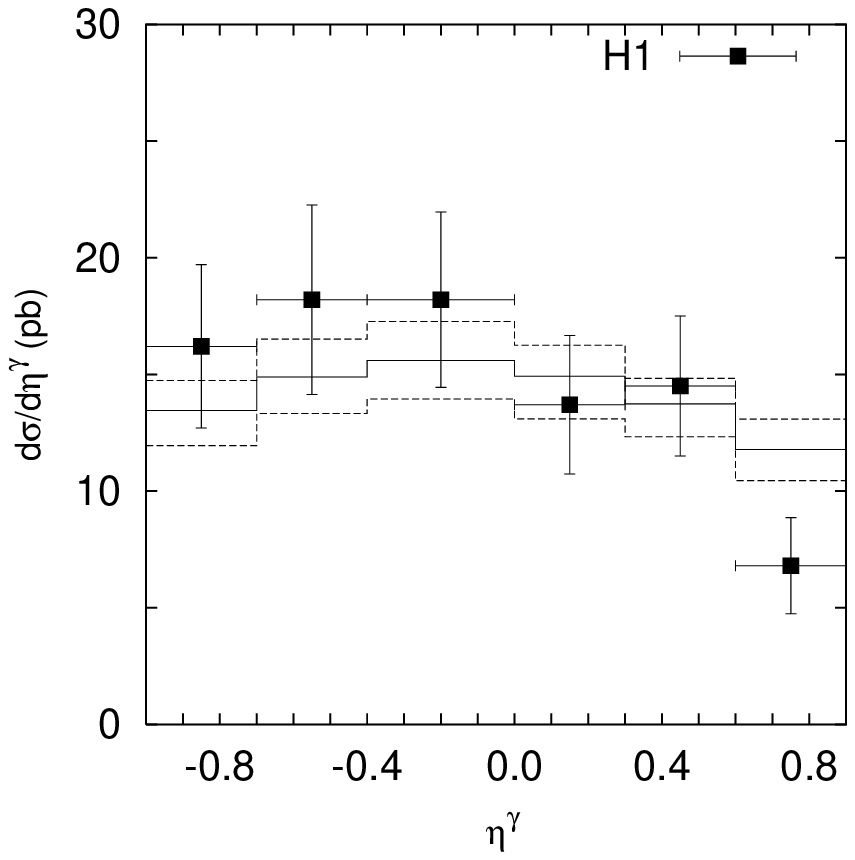}
\end{minipage}
\caption{
Same as Fig.~\ref{fig:h1cut}, but for $\gamma$+jet events with  
the additional jet cuts: 
$ - 1 < \eta^{\rm jet} < 2.3$ and $E_T^{\rm jet} > 4.5$ GeV.
}
\label{fig:h1cut2}
\end{figure}

The H1 experimental data in photoproduction~\cite{Aktas:2004uv} were also compared to the
$k_T$-factorization approach~\cite{Lipatov:2005tz}. 
Comparison with the $k_T$ factorization approach indicates somewhat better agreement, as
shown in Fig.~\ref{fig:h1cut} (see \cite{Lipatov:2005tz} for details). 
One can see that the measured distributions are reasonably well described except
the moderate $E_{T}^\gamma$ region  and in the 
pseudorapidity region $ - 0.4 \leq \eta^\gamma \leq 0.9$ only. For $ - 1 \leq \eta^\gamma \leq -0.4$
the $k_T$-factorization  predictions are mostly below the experimental
points.
The discrepancy between data and theory at negative $\eta^\gamma$ is
found to be relatively strong at low values of the initial photon
fractional momentum $y$. The effect of scale variations in transverse energy distributions
is rather large: the relative difference
between results for $\mu = E_T^\gamma$ and results for $\mu = E_T^\gamma/2$ or
$\mu = 2 E_T^\gamma$ is about 15\% within the $k_T$-factorization approach, 
which is due to missing higher order corrections. The scale dependence of the 
 NLO QCD calculations in the collinear factorization approach is below the 10\% level.
 
The individual contributions from the direct and resolved production
mechanisms to the total cross section in the $k_T$-factorization approach
is about 47\% and 53\%, respectively.  The contributions of single resolved processes
$$
  q (k_1) + g (k_2) \to \gamma (p_{\gamma}) + q (p'), 
$$
$$
  g (k_1) + q (k_2) \to \gamma (p_{\gamma}) + q (p'),
$$
$$
  q (k_1) + q (k_2) \to \gamma (p_{\gamma}) + g (p'). 
$$
account for 80\%, 14\% and 6\% respectively.

The transverse energy $E_T^\gamma$ and pseudorapidity $\eta^\gamma$ distributions 
for $\gamma$+jet events measured by H1 are compared to the  $k_T$-factorization predictions in Fig.~\ref{fig:h1cut2} (see 
also Ref.~\cite{Lipatov:2005tz}). 
In contrast to the inclusive case, one
can see that the $k_T$-factorization predictions are consistent with the data in most bins, although some discrepancies are present.
The theoretical results are lower than the experimental data at negative $\eta^\gamma$
and higher at positive $\eta^\gamma$.
The scale variation as it was described above changes the estimated cross sections by about 10\%. 
Note that such disagreement between predicted and measured
cross sections has also been observed for the NLO QCD calculations in the 
collinear factorization approach, 
see Fig.~\ref{fig:H1jets}.

\begin{figure}[ht]
\begin{minipage}[b]{0.48\linewidth}
\centering
\includegraphics[scale=0.7]{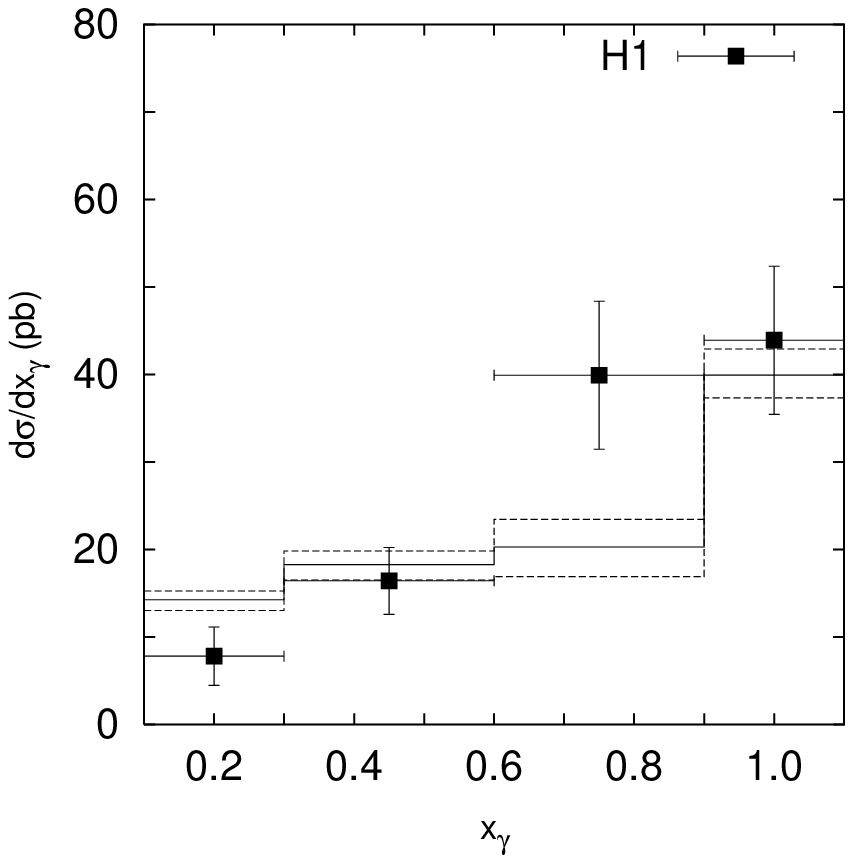}
% \caption{default}
% \label{fig:figure1}
\end{minipage}
\hspace{0.5cm}
\begin{minipage}[b]{0.48\linewidth}
\centering
\includegraphics[scale=0.7]{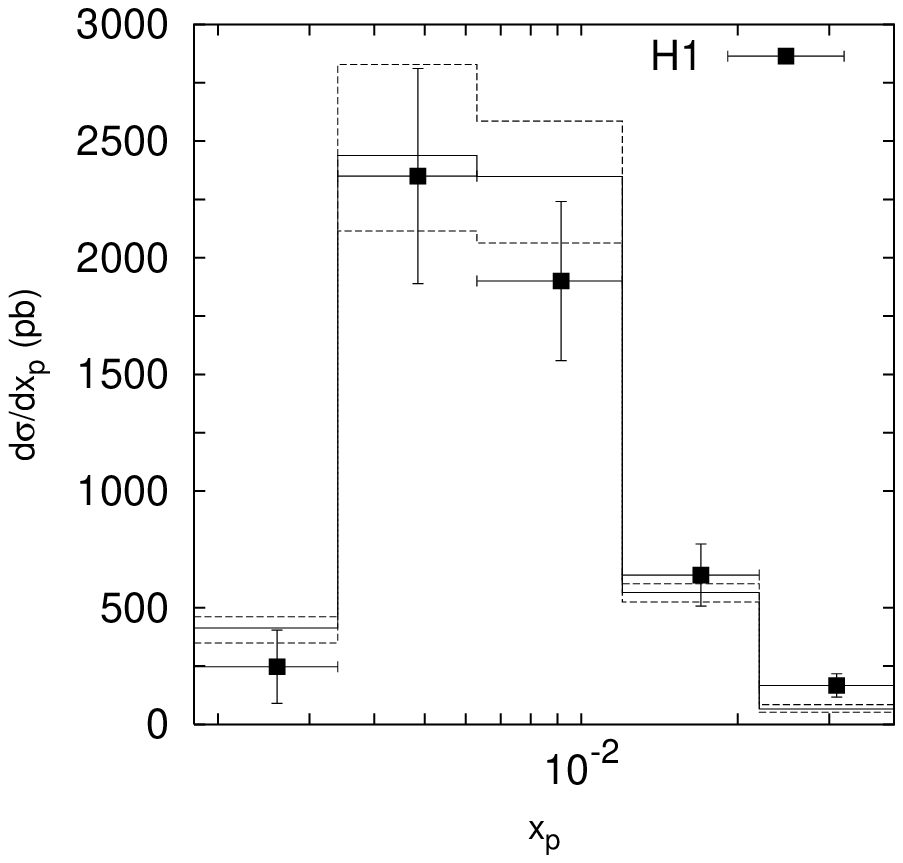}
\end{minipage}
\caption{
The differential  cross section $d\sigma/dx_{\gamma}$ and $d\sigma/dx_p$
 of prompt photon + jet  production calculated  at $5 < E_T^\gamma < 10$ GeV and $0.2 
< y < 0.7$ with an additional jet requirement $ - 1 < \eta^{\rm jet} < 2.3$ and $E_T^{\rm jet} > 4.5$ GeV.
}
\label{xgam}
\end{figure}

\noindent

Figure~\ref{xgam} shows the $x_{obs}^{\gamma}$ and  $x_{obs}^{p}$ 
distributions (see Eq.~\ref{x_gam})  measured by  H1.
One can see that $k_T$ factorization predictions  
reasonably well  agree with the experimental data. 
The NLO calculations~\cite{ejp:c21:303,pr:d64:14017} without corrections
for hadronization and multiple interactions give similar results.

\begin{figure}[ht]
\begin{minipage}[b]{0.48\linewidth}
\centering
\includegraphics[scale=0.5]{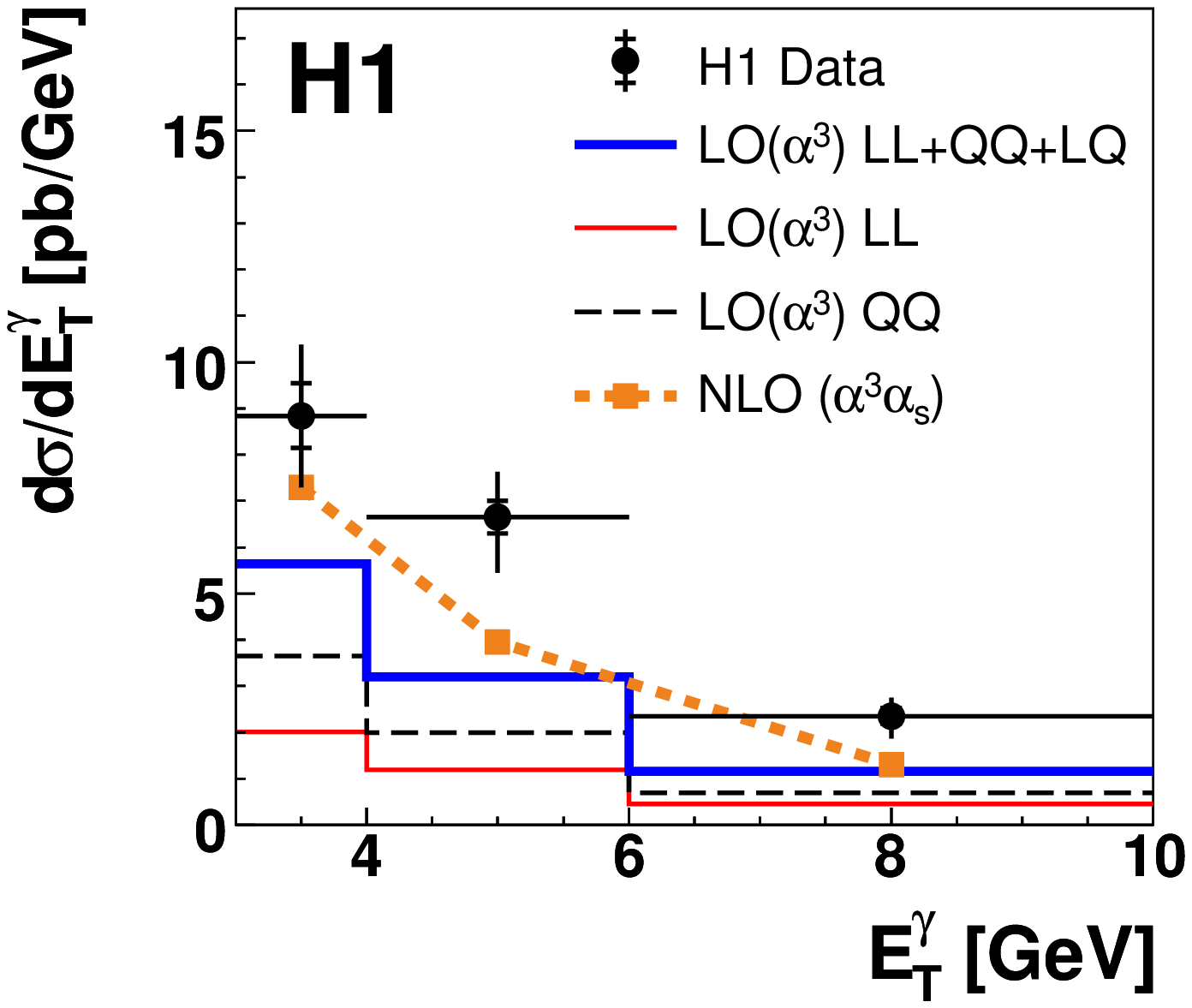}
% \caption{default}
% \label{fig:figure1}
\end{minipage}
\hspace{0.5cm}
\begin{minipage}[b]{0.48\linewidth}
\centering
\includegraphics[scale=0.5]{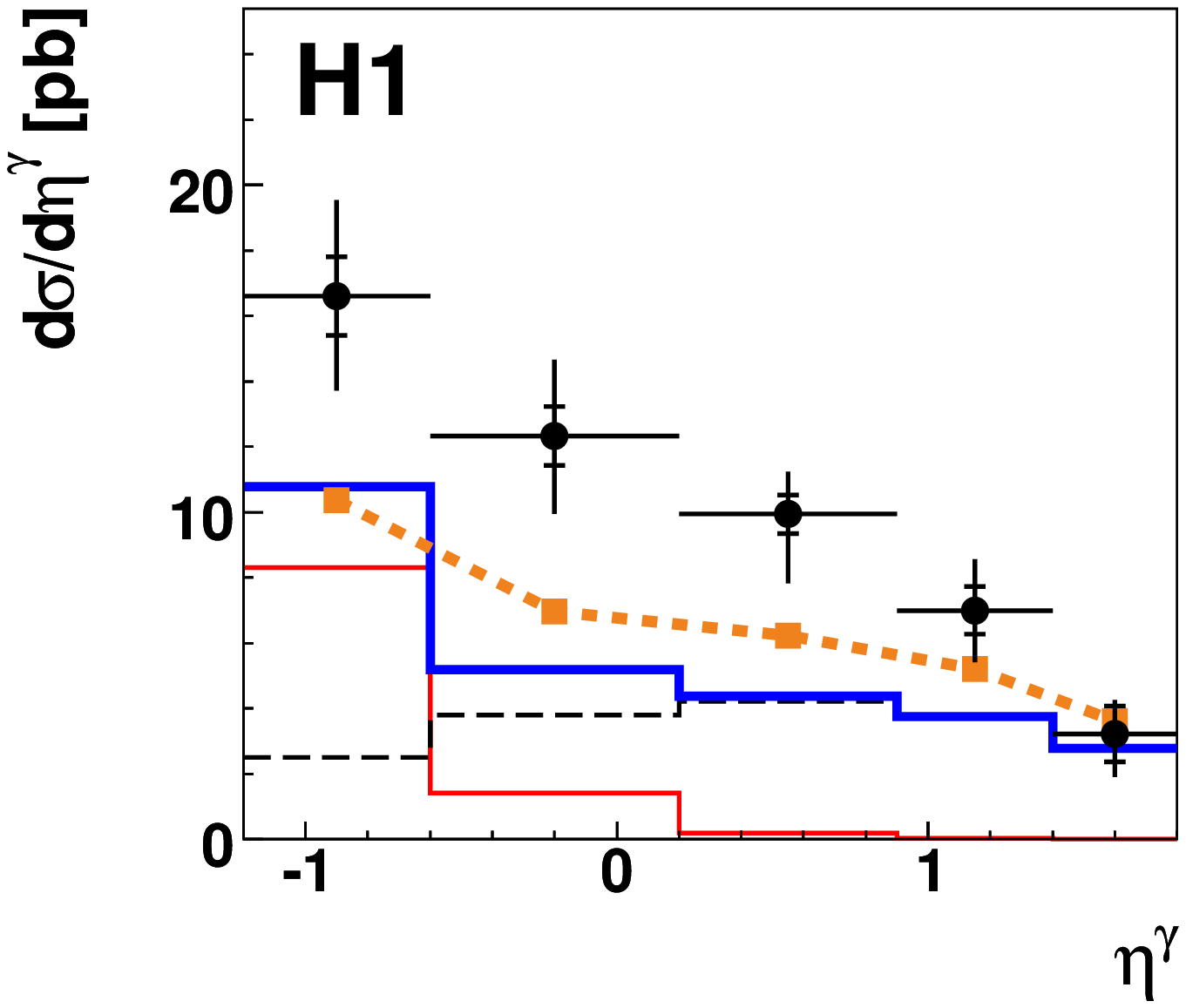}
%  \caption{default}
% \label{fig:figure2}
\end{minipage}
\caption{
The differential $\gamma$+jet cross sections
as functions of $E_{T}$ and $\eta$  of  the prompt
photon  in DIS.
The data are compared to LO and NLO calculations.}
\label{fig:h1}
\end{figure}

The H1 Collaboration~\cite{:2007eh} also has performed $\gamma$+jet measurements in DIS
for $Q^2>4$ GeV$^2$.
The NLO calculations~\cite{GehrmannDeRidder:2000ce},
which are only available for  $\gamma$+jet final state,
failed to describe normalization of the cross sections, although
the agreement in shape was found to be reasonable (Fig.~\ref{fig:h1}).
No $k_T$ factorization prediction available for DIS.

In summary, some differences with NLO QCD were observed in both photoproduction
and DIS. 
Differences at low $P_T^{\gamma}$ can be due to 
the treatment of the fragmentation contribution in NLO calculations.
Further, it would be interesting to see the effect of calculations beyond NLO QCD. 
The approach based on the $k_T$ factorization has better agreement with the data, but
such calculations have larger theoretical uncertainties.

\section{Comparison with Tevatron results}

Isolated photons in $p\bar{p}$ collisions at Tevatron have 
been measured recently by the CDF~\cite{Acosta:2002ya,Acosta:2004bg}
and D0~\cite{Abbott:2000,Abazov:2001af,Abazov:2005wc,Abazov:2008er} Collaborations.

\begin{figure}[!ht]
\centering
\includegraphics[scale=0.5]{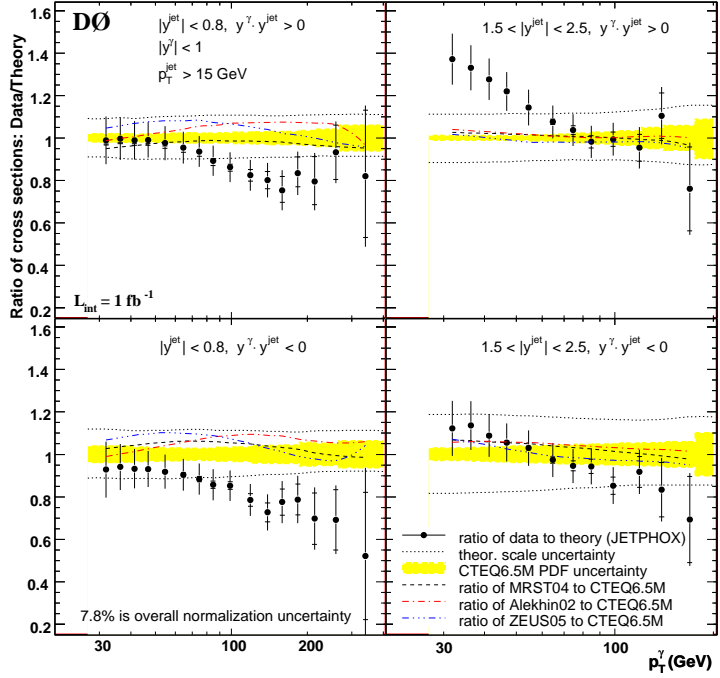}
\caption{
The ratios of the triple-differential cross section measured by D0 compared to
to the NLO QCD prediction using JETPHOX. See details in Ref.~{\protect \cite{Abazov:2008er}}.
}
\label{d0fig7}
\end{figure}

\begin{figure}[!ht]
\begin{minipage}[b]{0.48\linewidth}
\centering
\includegraphics[scale=0.7]{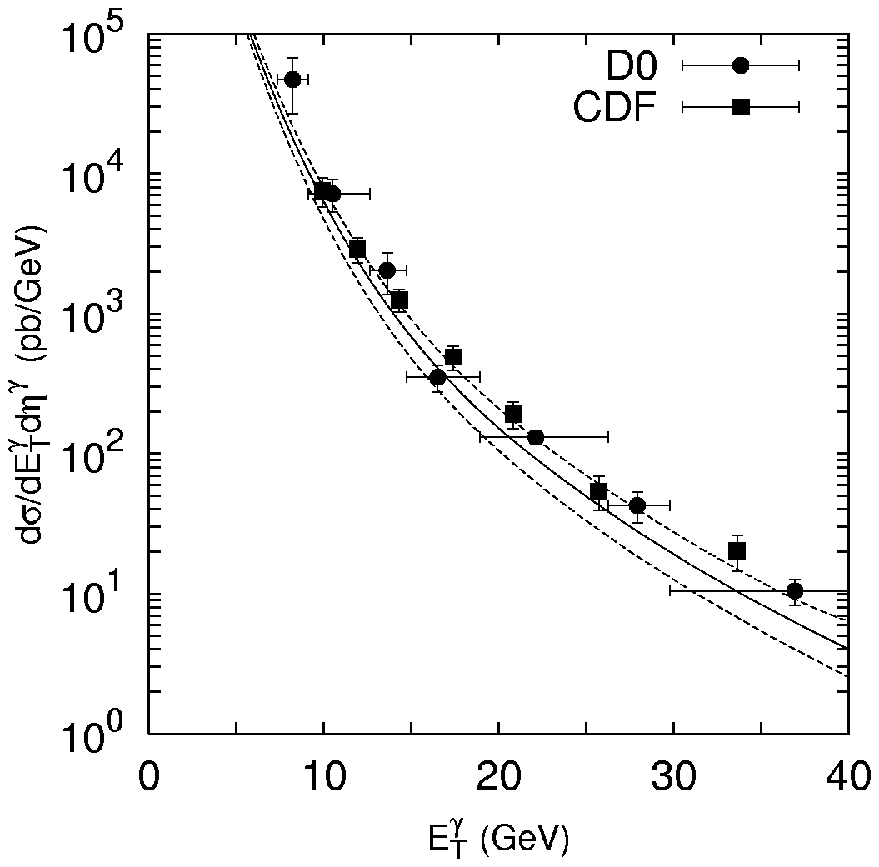}
% \caption{default}
% \label{fig:figure1}
\end{minipage}
\hspace{0.5cm}
\begin{minipage}[b]{0.48\linewidth}
\centering
\includegraphics[scale=0.7]{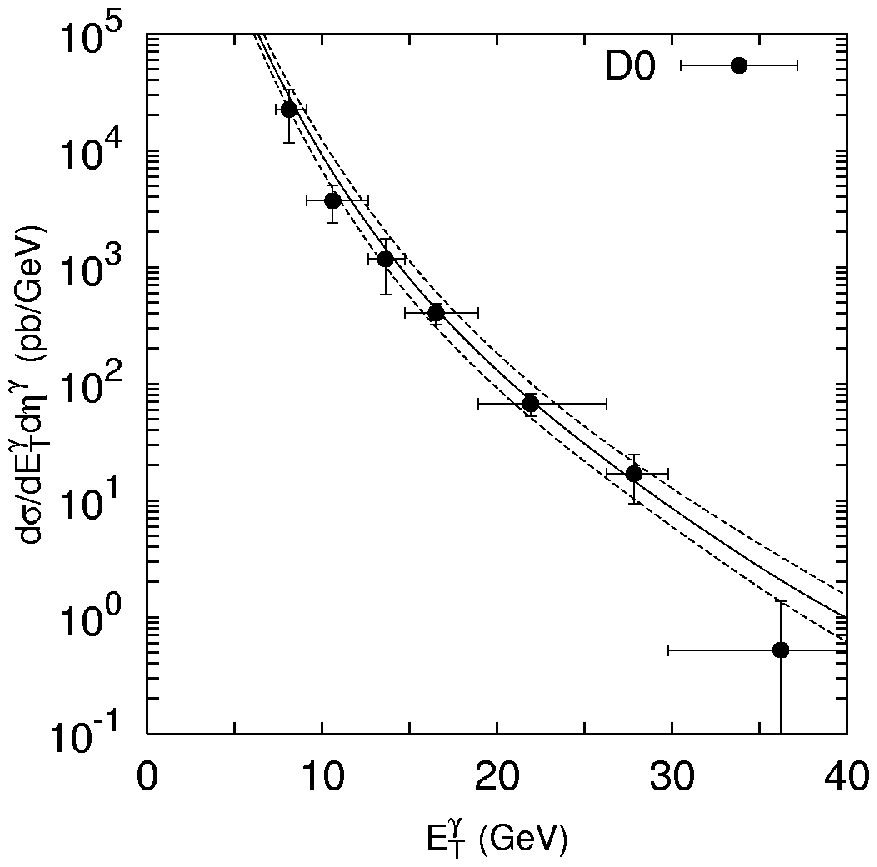}
\end{minipage}
\caption{
The double differential  cross section $d\sigma/dE_T^{\gamma}d\eta^\gamma$
 of inclusive prompt photon production at $\sqrt s = 630$ GeV and $|\eta^\gamma| <
0.9$ (left plot) and $1.6 < |\eta^\gamma| <  2.5$ (right panel). The solid line corresponds to the default scale $\mu = E_T^\gamma$ of the $k_T$ factorization predictions, whereas upper and lower dashed lines correspond to the  $\mu = E_T^{\gamma}/2$ and $\mu = 2E_T^{\gamma}$.
}
\label{tev1}
\end{figure}

\begin{figure}[!ht]
\begin{minipage}[b]{0.48\linewidth}
\centering
\includegraphics[scale=0.7]{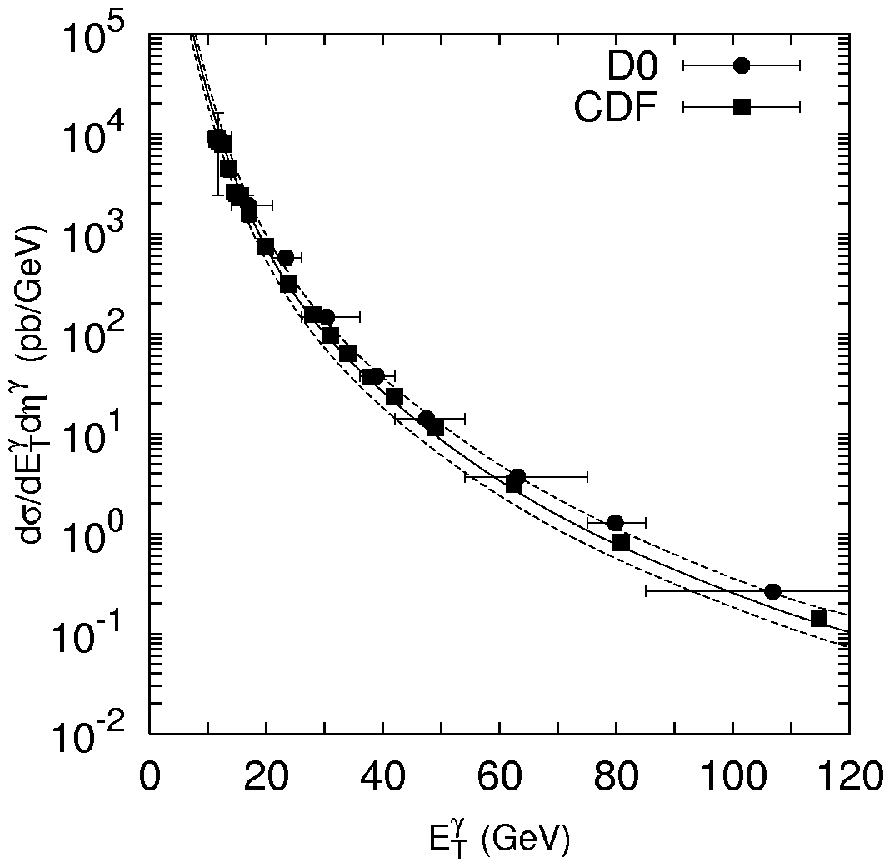}
% \caption{default}
% \label{fig:figure1}
\end{minipage}
\hspace{0.5cm}
\begin{minipage}[b]{0.48\linewidth}
\centering
\includegraphics[scale=0.7]{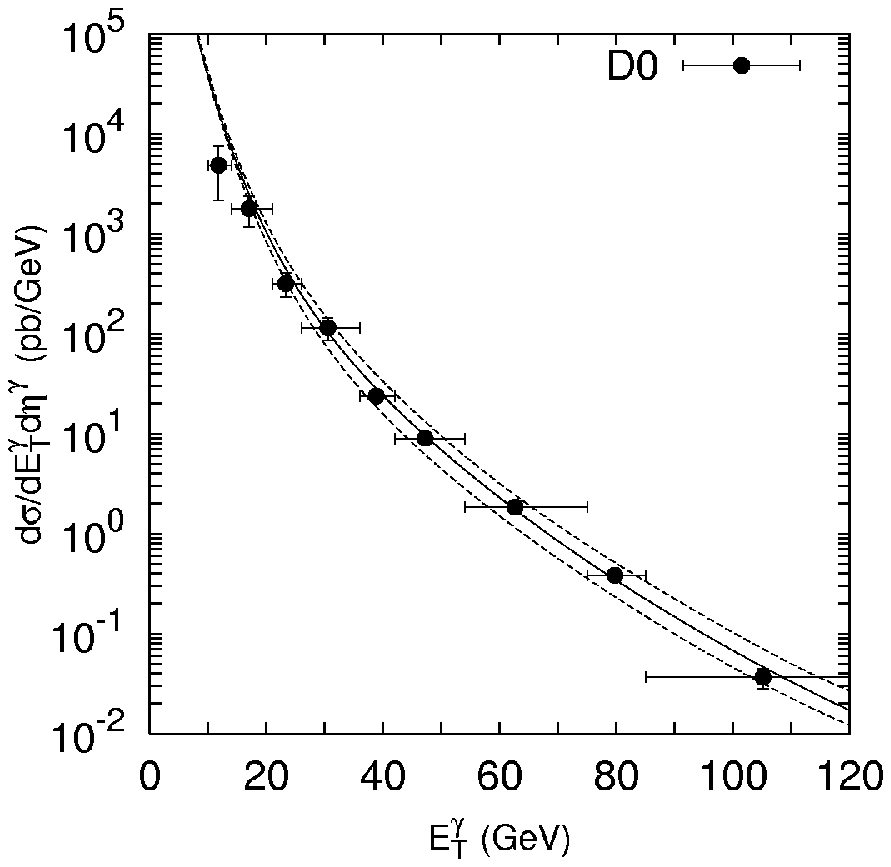}
\end{minipage}
\caption{
The double differential  cross section $d\sigma/dE_T^{\gamma}d\eta^\gamma$
 of inclusive prompt photon production at $\sqrt s = 1800$ GeV and $|\eta^\gamma| <
0.9$ (left plot) and $1.6 < |\eta^\gamma| <  2.5$ (right plot).
The solid line corresponds to the default scale $\mu = E_T^\gamma$, whereas upper and
lower dashed lines correspond to the  $\mu = E_T^{\gamma}/2$ and $\mu = 2E_T^{\gamma}$ for the
$k_T$ factorization calculations.
}
\label{tev2}
\end{figure}

Measurements of $p\,\bar{p} \to \gamma +{\rm{jet}}+X$ for
30 GeV \,$\leq p_T^\gamma \leq 300 $\,GeV
have very recently been published by D0\,\cite{Abazov:2008er}.
The comparison to theory is done separately for different regions in rapidity
of the photon and the jet.
The NLO partonic Monte Carlo program JETPHOX~\cite{Aurenche:2006vj,phox}
was used to compare the data to theory at next-to-leading order.
It was shown that the NLO calculations are not sufficient to describe
the shape of $P_T^{\gamma}$ distributions in different rapidity regions, as can be seen
in Figure~\ref{d0fig7}. At present, the comparison with the
$k_T$-factorization prediction is in progress.

Differences with the collinear factorization approach have been seen previously as well.
Both CDF~\cite{Acosta:2002ya} and D0~\cite{Abazov:2001af}
cross sections were found to be above\footnote{For D0, the difference was mainly 
concentrated in the central rapidity  
region.} NLO predictions  at low $P_T^{\gamma}$.
However, RHIC has also measured prompt photon production in $pp$ collisions at 
$\sqrt{s}=200$ GeV and found good agreement with NLO theory in the 
collinear factorization approach~\cite{Adler:2006yt,Heinrich:2007mx}.

The same data were compared to the $k_T$ factorization approach in \cite{Lipatov:2005wk}.
Figures~\ref{tev1} and ~\ref{tev2} 
show the 
CDF~\cite{Acosta:2002ya} and D0~\cite{Abazov:2001af} measurements 
for the $d\sigma/dE_T^\gamma d \eta^\gamma$ cross sections calculated at $\sqrt s = 630$ and 1800 GeV
in central and forward kinematic regions together with the $k_T$ factorization
predictions.  
One can see that theoretical predictions agree
with the experimental data within the scale uncertainties.
However, the results of the calculation with the default scale
tend to underestimate the data in the central kinematic region
and agree with the D0 data in the forward $\eta^\gamma$ region.
The collinear NLO QCD calculations give
a similar description of the data: %the results of measurement are higher
% than the NLO prediction at low $E_T^\gamma$ in the central $\eta^\gamma$ range
% but agree at all $E_T^\gamma$ in the forward pseudo-rapidity region.
generally there is a residual negative slope in the ratio of the data over the prediction as a function of $E_T^\gamma$.
The scale dependence of the $k_T$ factorization results is
rather large ($20-30\%$), due to the fact that 
these are leading order calculations.
% and  compatible with that for the collinear NLO calculations.

The double differential cross sections $d\sigma/dE_T^\gamma d \eta^\gamma$
are usually the most difficult observables to describe using QCD predictions.
Yet, as 
it can be seen from Fig.~\ref{tev2},
the $k_T-$factorization predictions agree well with 
D0~\cite{Abbott:2000} and CDF~\cite{Acosta:2002ya} data
both in shape and normalization.
There are only rather
small overestimations of the data at low $E_T^\gamma$ values in
Figs.~\ref{tev2} in the forward region. Again, the scale dependence of our calculations
is about 20--30\%.
The theoretical uncertainties of the collinear NLO predictions are
smaller (about 6\%~\cite{Abbott:2000}), which is to be expected as inclusion of 
higher order terms reduces the scale uncertainty.

One can conclude that the results of calculations in the
$k_T-$factorization approach  in general
agree well with Tevatron experimental data, within a large scale uncertainty.

\section{Prompt photons at LHC}

\begin{wrapfigure}{r}{0.35\columnwidth}
\centerline{\includegraphics[width=0.35\columnwidth]{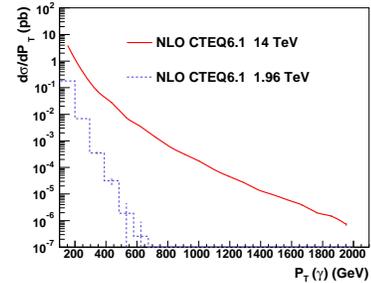}}
\caption{\label{fig:tt}
The $P_T^{\gamma}$ cross section for $\gamma$+jet events
predicted by NLO QCD for the Tevatron and the LHC kinematic range.
}
\end{wrapfigure}

The direct photon production at LHC 
has significantly higher cross sections compared to the ones measured at
Tevatron and HERA. 
The prompt-photon cross section at LHC is more than a factor of hundred higher
than that at Tevatron and a factor of $10^{5}$ larger than that for 
photoproduction at HERA,
assuming a similar kinematic range ($\mid \eta^{\gamma} \mid < 2$),
as shown in Fig.~\ref{fig:tt}.
This will allow to explore the TeV energy scale already in a few years of data taking.

\begin{figure}[ht]
  \begin{center}
   \includegraphics[scale=0.55]{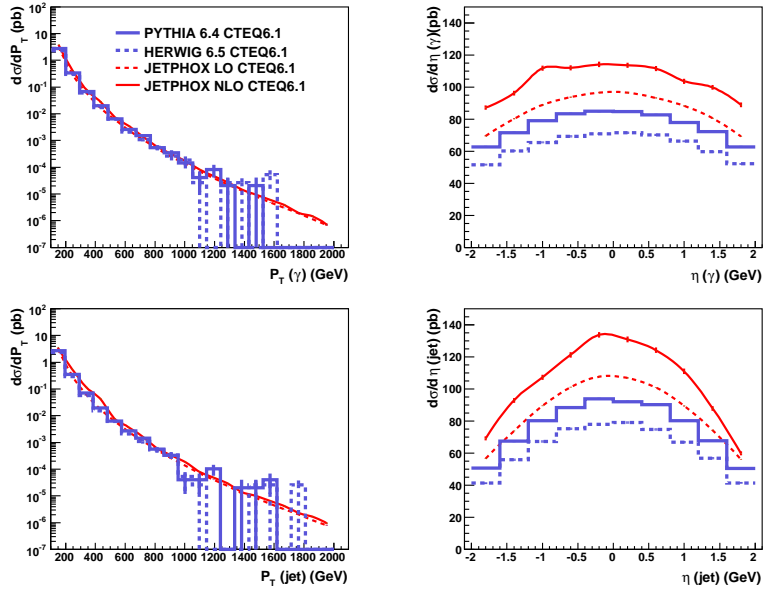}
   \caption{Comparisons of Monte Carlo models with LO and NLO calculations as implemented in JETPHOX.
      }
    \label{fig:mc}
  \end{center}
\end{figure}

Figure~\ref{fig:mc} shows the comparison between PYTHIA and HERWIG Monte Carlo models and JETPHOX
LO and NLO calculations. 
The cross sections for $\gamma+jet$ events were calculated for $\mid \eta^{\gamma} \mid < 2$, 
$P_T^{\gamma}>100$ GeV and $P_T^{jet}>105$ GeV. 
The cuts on the transverse momenta are asymmetric to avoid instabilities in the NLO calculations.  
An  isolation requirement
$E_{T}^{\gamma}>0.9\, E_{T}^{tot}$ was imposed, where
$E_{T}^{tot}$ is the total energy of the jet which contains prompt photon.
Jets were reconstructed with
the longitudinally-invariant
$k_{T}$ algorithm in inclusive mode \cite{pr:d48:3160,*np:406:187}.

The NLO QCD calculation is 30--40\% higher than     
that predicted by PYTHIA. On the other hand, PYTHIA is $20\%$ above HERWIG.  
It is interesting to observe that the level of discrepancy between PYTHIA and
HERWIG is about the same as that observed at HERA at much lower transverse momenta (for example see Fig.~\ref{fig:zeus}).
However, there is no significant difference between NLO and PYTHIA at $P_T^{\gamma}>10$ GeV for $ep$, while 
at the LHC energy range the difference between NLO and PYTHIA is rather significant.
Certainly, the overall normalization of Monte Carlo programs like   PYTHIA or HERWIG
has to be adjusted, as these programs cannot account for contributions from 
loop corrections at higher orders.

\begin{figure}[ht]
\begin{minipage}[b]{0.5\linewidth}
\centering
\includegraphics[scale=0.4]{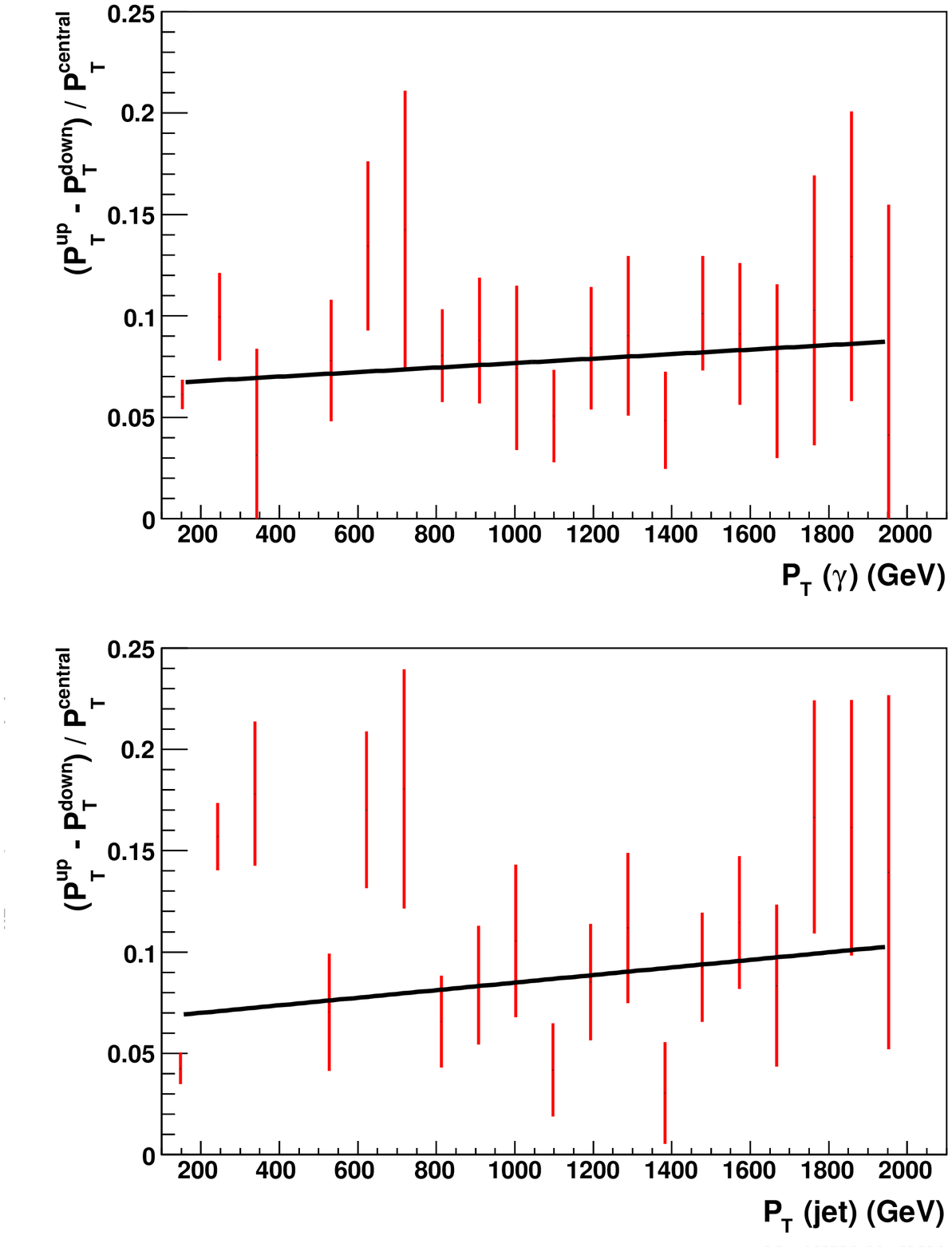}
\caption{Relative difference between the cross section estimated with 
$\mu =0.5$ ($P_T^{up}$) and $\mu=2$ ($P_T^{down}$) as a function of $P_T$ for gamma and jet. The line represents
a linear fit. }
\label{fig:figure1}
\end{minipage}
\hspace{0.5cm}
\begin{minipage}[b]{0.5\linewidth}
\centering
\includegraphics[scale=0.4]{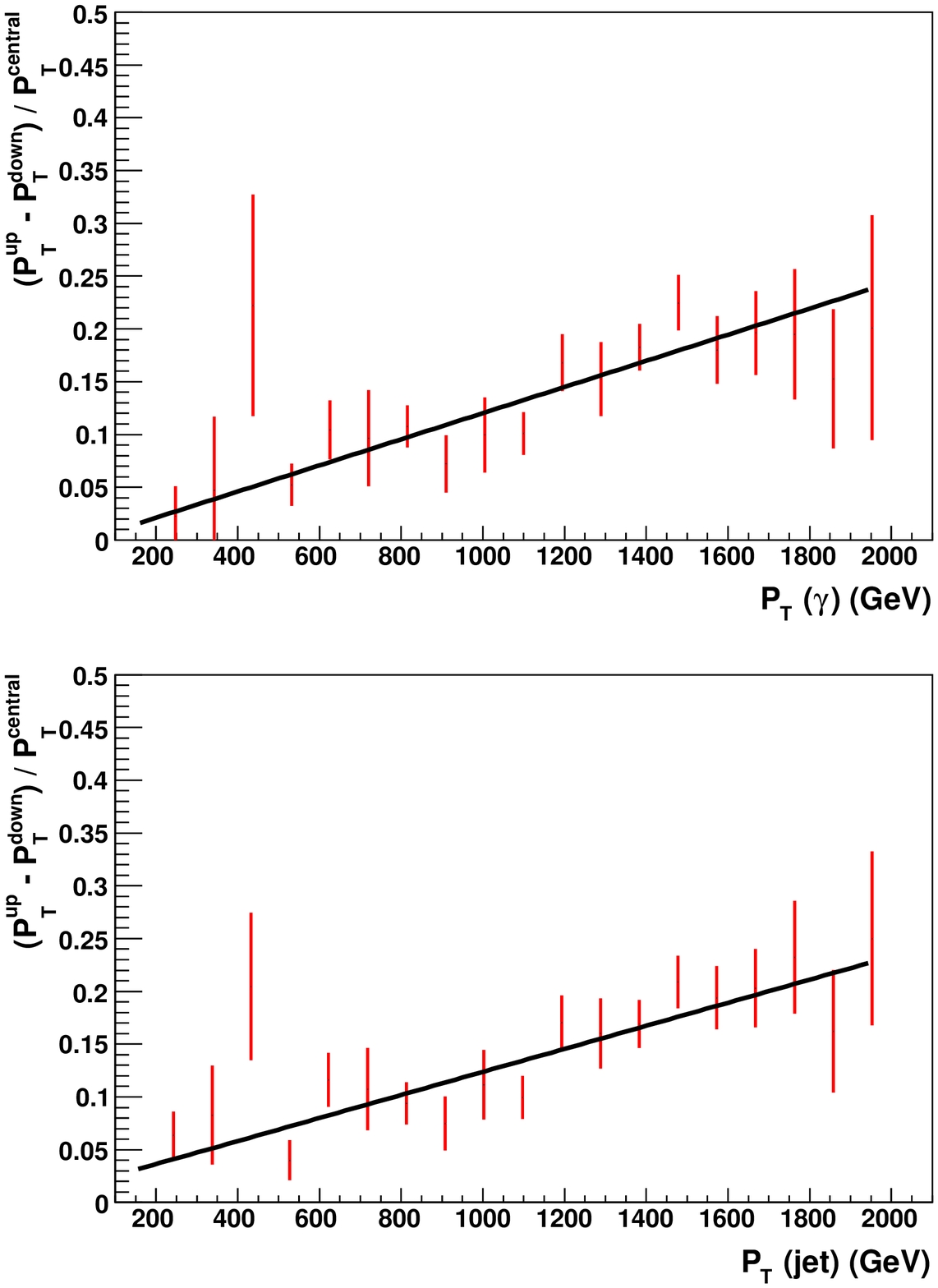}
\caption{
Relative difference between the cross section estimated with  
CTEQ6.1M set=-15 ($P_T^{up}$) and set =15  ($P_T^{down}$) as a function of $P_T$ for gamma and jet. The line represents 
a linear fit. 
}
\label{fig:figure2}
\end{minipage}
\end{figure}

Scale uncertainties were estimated by changing the renormalization and factorization scales in the range $0.5<\mu_f, \mu_R < 2$.
The relative difference between predicted cross sections is shown in Fig.~\ref{fig:figure1}.
To make quantitative statements on scale uncertainties with the present level of statistical 
errors in calculations using JETPHOX,
a linear fit was performed to determine the trend
of the relative differences with increase of $P_T^{\gamma}$. 
As it can be seen, the scale uncertainty is about $10\%$ and slowly
increases with $P_T^{\gamma}$ .

To estimate the uncertainty associated with the gluon density, the calculations have been
performed using two CTEQ6.1M sets (15 and 30) which correspond to two extremes in the gluon density
at large $x$ \cite{steve}.         
Fig.~\ref{fig:figure2} shows the relative difference between those two sets as a function of $P_T^{\gamma}$.
It is seen that the gluon uncertainty is almost a factor of two larger compared to the scale
uncertainty estimated above. No statistically significant difference has been observed between 
the cross sections calculated using CTEQ6.1M and MRST04.
This is not totally surprising as both sets have similar input data for the global fit analysis.

\begin{figure}[ht]
\begin{minipage}[b]{0.48\linewidth}
\centering
\includegraphics[scale=0.7]{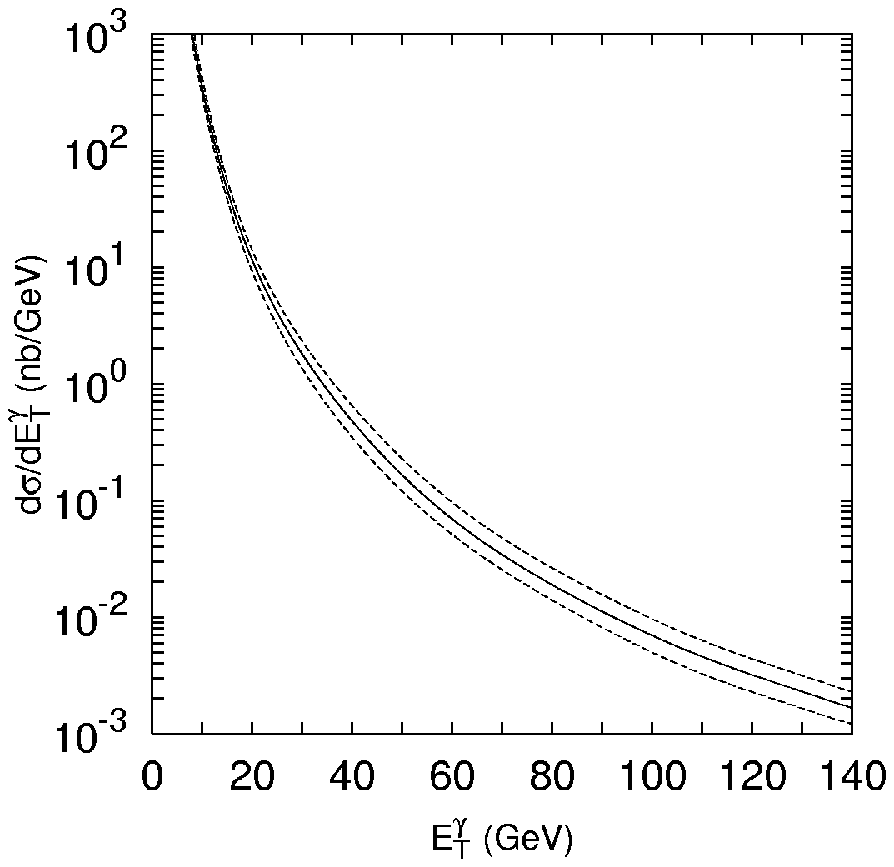}
\end{minipage}
\hspace{0.5cm}
\begin{minipage}[b]{0.48\linewidth}
\centering
\includegraphics[scale=0.7]{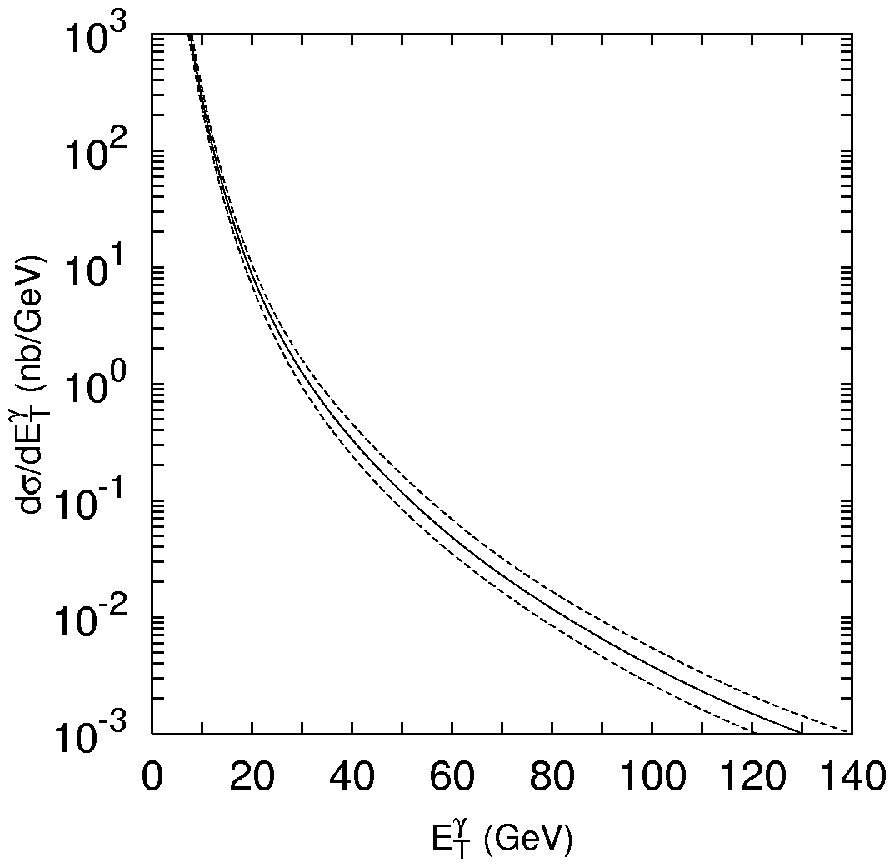}
\end{minipage}
\caption{Left plot:
The  $k_T$ factorization predictions for differential cross sections $d\sigma/dE_T^{\gamma}d\eta^\gamma$
at $\sqrt s = 14$ TeV GeV and $|\eta^\gamma| <
2.5$ (left plot); at $2.5 < \eta^\gamma < 4.0$ (right plot).
The solid line corresponds to the default scale $\mu = E_T^\gamma$,
whereas upper and lower dashed lines correspond to the  $\mu = E_T^{\gamma}/2$ and $\mu = 2E_T^{\gamma}$.
}
\label{lhc_kt}
\end{figure}

The predictions for the $k_T$ factorization approach were obtained
for a wider pseudorapidity range, 
for both
central and forward pseudo-rapidities $\eta^\gamma$.
As a representative example,
we will define the central and forward kinematic regions
by the requirements $|\eta^\gamma| < 2.5$ and
$2.5 < |\eta^\gamma| < 4$, respectively.
The transverse energy $E_T^\gamma$
distributions of the inclusive prompt photon production in different
$\eta^\gamma$ ranges at $\sqrt s = 14$ TeV
are shown in Figs.~\ref{lhc_kt}.  One can see that variation in scale $\mu$
changes the estimated cross sections by about 20--30\%.
However, as it was already discussed above, there are
additional theoretical uncertainties due to
the non-collinear parton evolution, and these uncertainties are not
well studied up to this time. Also the extrapolation
of the available parton distribution to the region
of lower $x$ is a special problem at the LHC
energies. In particular, one of the problem
is connected with the correct treatment of saturation effects
in small $x$ region.
Therefore,  more work needs to be done
until these uncertainties will be reduced.

Thus, the calculation based on the $k_T$ factorization approach shows a
larger scale uncertainty
compared to the collinear factorization approach: for $P_T^{\gamma} \sim 100$ GeV, 
the overall uncertainty for the
NLO calculations is expected to be around $10\%$, while it reaches 
20--30\% for the $k_T$-factorization calculations for the same $P_T^{\gamma}$ range, 
due to the fact that the latter are at leading order in $\alpha_s$. 
As the residual scale dependence of missing higher order 
terms resides in logarithms  involving ratios of $P_T^2$ and scales $\mu^2$, 
the effect becomes more dramatic at the LHC energy.

%This conclusion is similar to that obtained by analyzing HERA and Tevatron data.  
%For the LHC, 
%the difference in the size of the uncertainties between the two factorization approaches seems to be more dramatic. 

\section{Summary}

In this review, we have attempted to summarize recent progress
in the description of prompt photon production at HERA, the Tevatron and the LHC.
At HERA, some differences with NLO were observed in both photoproduction
and DIS. The deficiencies at low $P_T^{\gamma}$
values may  
indicate that non-perturbative effects 
at small $P_T^{\gamma}$ play a non-negligible role.
Also, one  should expect that adding high-order corrections
to the collinear-factorization approach should improve the description.
Similar conclusions  can be drawn 
for the Tevatron data which, as in the HERA case, has differences 
with NLO in the lowest $P_T^{\gamma}$ region.
Recently, significant differences with NLO were observed by the Tevatron for the shapes of $P_T^{\gamma}$ 
distributions differential in $\eta^{\gamma}$. 
On the other hand, RHIC observes good agreement with NLO QCD. 
Considering the fact that RHIC uses a photon isolation method which is 
different from the usual cone isolation, the differences mentioned above 
may also have to do with isolation criteria acting differently in a partonic 
calculation than in the full hadronic environment of the experiment.

An alternative approach based on the $k_T$ factorization generally improves the 
description of the HERA and the Tevatron data, but it has larger theoretical uncertainties.
As for NLO, high-order corrections
to the $k_T$-factorization approach should improve the description of the data.
The applicability of the $k_T$ factorization to the LHC data will be tested with the arrival of 
the first LHC data, but it is already evident that
significant theoretical uncertainties are expected for the description of prompt-photon
cross sections at LHC. Using the the collinear factorization approach,
uncertainties of NLO calculations are expected to be 10--20\% at about 1 TeV photon transverse momenta,
and significantly larger for the $k_T$-factorization calculations. These uncertainties have to be reduced 
in the future for detailed comparison of the LHC data with the QCD predictions.       

In all cases, Monte Carlo predictions fail to describe  prompt-photon
cross sections,  both in shape and normalization. Generally, HERWIG is significantly
below PYTHIA. This could have a direct impact on the future LHC               
measurements, in particular for exotic searches which often rely on Monte Carlo
predictions for estimations of rates for background events.

This work supported in part by the U.S. Department of Energy, Division of High Energy Physics, under Contract DE-AC02-06CH11357. 

%------------------------------------------------------------------------------
%       Bibliography
%------------------------------------------------------------------------------
% \bibliographystyle{heralhc} 
\newpage 
\bibliographystyle{h-physrev3.bst}
{\raggedright
\bibliography{heralhc}
}
\end{document}